\documentclass{pasa}%
\usepackage{bigdelim}
\usepackage{graphicx}
\usepackage{lscape}
\usepackage{longtable}
\usepackage{todonotes}
\usepackage[figuresright]{rotating}
\usepackage{threeparttable}
\usepackage{hyperref}
\usepackage{multirow}
\usepackage{subcaption}
\usepackage{cprotect}
\usepackage{chngcntr}
\usepackage{amssymb}
\usepackage{upgreek}

\usepackage{etoolbox}

\def\nar{\ref@jnl{New A Rev.}}          % New Astronomy Review

\title[POGS II: All-Sky Results]{The POlarised GLEAM Survey (POGS) II: Results from an All-Sky Rotation Measure Synthesis Survey at Long Wavelengths}

\author[Riseley et al.]{C.~J.~Riseley$^{1,2,3}$\thanks{Corresponding author email: \url{christopher.riseley@unibo.it}}, \hspace{0.001cm} 
T.~J.~Galvin$^{3}$, C.~Sobey$^{3}$, T.~Vernstrom$^{3}$, S.~V.~White$^{4,5}$, X.~Zhang$^{3}$, 
B.~M.~Gaensler$^{6}$, G.~Heald$^{3}$, 
C.~S.~Anderson$^{7,3,8}$, T.~M.~O.~Franzen$^{9}$, P.~J.~Hancock$^{5}$, N.~Hurley-Walker$^{5}$, E.~Lenc$^{8}$, C.~L.~Van~Eck$^{6}$ 
\affil{$^1$ Dipartimento di Fisica e Astronomia, Universit\`a degli Studi di Bologna, via P. Gobetti 93/2, 40129 Bologna, Italy}
\affil{$^2$ INAF -- Istituto di Radioastronomia, via P. Gobetti 101, 40129 Bologna, Italy}
\affil{$^3$ CSIRO Astronomy and Space Science, PO Box 1130, Bentley, WA 6102, Australia}
\affil{$^4$ Department of Physics and Electronics, Rhodes University, PO Box 94, Grahamstown, 6140, South Africa}
\affil{$^5$ International Centre for Radio Astronomy Research, Curtin University, Bentley, WA 6102, Australia}
\affil{$^6$ Dunlap Institute for Astronomy and Astrophysics, 50 St. George St, University of Toronto, ON M5S 3H4, Canada}
\affil{$^7$ National Radio Astronomy Observatory, 1003 Lopezville Road, Socorro, NM 87801, USA}
\affil{$^8$ CSIRO Astronomy and Space Science, PO Box 76, Epping, NSW 1710, Australia}
\affil{$^9$ ASTRON: the Netherlands Institute for Radio Astronomy, Postbus 2, 7990 AA, Dwingeloo, The Netherlands}
}%0

\jid{PASA}
\doi{10.1017/pas.\the\year.xxx}
\jyear{\the\year}

\usepackage{aas_macros}
\usepackage{hyperref} 
\hypersetup{colorlinks,citecolor=blue,linkcolor=blue,urlcolor=blue}
\defcitealias{Riseley2018}{Paper I}

\begin{document}

\begin{frontmatter}
\maketitle

\begin{abstract}
The low-frequency linearly-polarised radio source population is largely unexplored. However, a renaissance in low-frequency polarimetry has been enabled by pathfinder and precursor instruments for the Square Kilometre Array. In this second paper from the POlarised GaLactic and Extragalactic All-Sky Murchison Widefield Array (MWA) Survey --- the POlarised GLEAM Survey, or POGS --- we present the results from our all-sky MWA Phase I Faraday Rotation Measure survey. Our survey covers nearly the entire Southern sky in the Declination range $-82\degree$ to $+30\degree$ at a resolution between around three and seven arcminutes (depending on Declination) using data in the frequency range 169$-$231\,MHz. We have performed two targeted searches: the first covering 25,489 square degrees of sky, searching for extragalactic polarised sources; the second covering the entire sky South of Declination $+30\degree$, searching for known pulsars. We detect a total of 517 sources with 200\,MHz linearly-polarised flux densities between 9.9\,mJy and 1.7\,Jy, of which 33 are known radio pulsars. All sources in our catalogues have Faraday rotation measures in the range $-328.07$~rad~m$^{-2}$ to $+279.62$~rad~m$^{-2}$. The Faraday rotation measures are broadly consistent with results from higher-frequency surveys, but with typically more than an order of magnitude improvement in the precision, highlighting the power of low-frequency polarisation surveys to accurately study Galactic and extragalactic magnetic fields. We discuss the properties of our extragalactic and known-pulsar source population, how the sky distribution relates to Galactic features, and identify a handful of new pulsar candidates among our nominally extragalactic source population.
\end{abstract}

\begin{keywords}
polarisation -- surveys -- radio continuum: general -- galaxies: active -- pulsars
\end{keywords}
\end{frontmatter}

\section{INTRODUCTION} \label{sec:intro}
The construction of an all-sky grid of polarised sources with which to probe the large-scale magnetised Universe is one of the high-priority goals of many surveys with next-generation radio telescopes. For example, the POlarisation Sky Survey of the Universe's Magnetism \citep[POSSUM;][]{Gaensler2010} project with the Australian Square Kilometre Array Pathfinder \citep[ASKAP;][]{Johnston2007} is predicted to detect up to a million polarised sources across the Southern sky (South of Declination around $+30\degree$). Such all-sky grids can be used as statistical probes of cosmic magnetism, and a number of large-$N$ studies have already been performed using previous state-of-the-art polarisation catalogues to probe both the Galactic magnetised foreground \citep[e.g.][]{Oppermann2012,Oppermann2015,Hutsch2020} and the extragalactic magnetised Universe \citep[e.g.][]{Vernstrom2019}.

However, these previous all-sky polarisation surveys -- principally the catalogue of \citet{Taylor2009}, produced from the NRAO VLA Sky Survey \citep[NVSS;][]{Condon1998} -- contain large systematic uncertainties, due to their poor frequency sampling \citep[e.g.][]{VanEck2011,Ma2019a}. However, next-generation surveys (such as POSSUM and the Karl G. Jansky Very Large Array Sky Survey, VLASS; \citealt{Lacy2020}) possess large fractional bandwidths that are finely-sampled in frequency, which mitigates many of the systematic uncertainties of the \citet{Taylor2009} catalogue.

Toward longer wavelengths, with the advent of next-generation interferometers such as the LOw-Frequency ARray \citep[LOFAR;][]{VanHaarlem2013} and the Murchison Widefield Array \citep[MWA;][]{Tingay2013}, low-frequency polarimetry has experienced a renaissance. Historically, such studies were extremely challenging, due to the poor sensitivity of previous instruments as well as calibration difficulties (related often to ionospheric effects). However, recent advances in calibration and imaging software, such as the MWA's Real-Time System \citep[RTS;][]{Mitchell2008} as well as techniques for correcting for instrumental leakage and ionospheric effects \citep[e.g.][]{Lenc2017,Lenc2018,Mevius2018} have enabled rapid progress in the field of low-frequency polarimetry. Another critical element that has enabled this renaissance is the availability of high-performance computing resources in the SKA precursor era. 

Indeed, in recent years the low-frequency polarised sky has been explored in greater detail than ever before, with detections of large-scale diffuse Galactic foreground along many lines of sight \citep[e.g.][]{Jelic2015,Lenc2016,VanEck2017,VanEck2019} as well as many surveys that have begun to study the polarised extragalactic source population \citep[e.g.][]{Bernardi2013,Mulcahy2014,VanEck2017,Lenc2018,Neld2018,OSullivan2019,OSullivan2020,Stuardi2020,Cantwell2020}.

\subsection{Faraday rotation}
Following, for example, \citet{Sokoloff1998}, the complex narrowband linear polarization $P$ can be expressed as:
\begin{equation}\label{eq:poldef}
    P = Q + iU = \Pi I{\rm{e}}^{2i\chi}
\end{equation}
where $I$, $Q$ and $U$ are the measured Stokes parameters, $\Pi$ is the fractional polarisation and $\chi$ is the polarisation angle. 

When a linearly-polarised radio wave encounters magnetised thermal plasma with some magnetic field component along the line-of-sight (LOS) from a given source to an observer, the plane of polarised emission rotates. This is known as Faraday rotation, and the observed frequency-dependent polarisation angle, $\chi(\lambda^2)$, is rotated according to:
\begin{equation}\label{eq:chi}
    \chi(\lambda^2) = \chi_0 + {\rm{RM}} \lambda^2
\end{equation}
where $\lambda$ is the observing wavelength and $\chi_0$ is the intrinsic source polarisation angle. The rotation measure (RM, in rad m$^{-2}$) is defined as:
\begin{equation}\label{eq:phi}
    {\rm{RM}} = 0.81 \int_{\ell}^0 n_{\rm{e}} B_{\|}\cdot {\rm{d}}{\ell}
\end{equation}
where $\ell$ is the distance to the source of polarised emission (in parsecs), $n_{\rm{e}}$ is the number density of free electrons (in cm$^{-3}$) and $B_{\|}$ is the magnetic field strength component along the LOS (in $\upmu$G). The accepted sign convention is that positive RM indicates a magnetic field oriented toward the observer, and negative RM indicates a magnetic field oriented away from the observer.

In regions where linearly polarised emission and Faraday rotation are co-located, interference effects along the LOS lead to more complex observed behaviour in $P(\lambda^2)$. Such behaviour carries a lot of physical information \citep[e.g.][]{Sokoloff1998} but is beyond the scope of our low-frequency polarimetry work. As quantified later in  Section~\ref{sec:data}, we are only sensitive to polarised emission that lies along the LOS through a medium that is purely Faraday-rotating \citep[i.e. does not also emit any polarised synchrotron emission itself, e.g.][]{Heald2009}.

\subsection{Surveys with the Phase I MWA}
The GaLactic and Extragalactic All-sky MWA survey \citep[GLEAM;][]{Wayth2015} covers the entire sky South of Declination $+30\degree$. The GLEAM Extragalactic Catalogue \citep{HurleyWalker2017} covers 24,831 square degrees of sky below this Declination and at Galactic latitudes $|b| \geq 10\degree$. In \citet{Riseley2018} (hereafter \citetalias{Riseley2018}) we applied recent technical advances in low-frequency polarisation calibration, as well as source-finding and verification techniques, to begin probing the low-frequency linearly-polarised source population across a wide area of the Southern sky. 

\citetalias{Riseley2018} presented the first results from the POlarised GLEAM Survey (POGS). We applied these novel data processing techniques to a sub-set of GLEAM data covering approximately 6400 square degrees of sky in the frequency range 200$-$231~MHz. This region covered 24h in Right Ascension and $20\degree$ in Declination centered on Declination~$-27\degree$. We detected 81 sources with polarised flux densities in excess of 18~mJy at 216\,MHz. In this paper, we present the full results from POGS, covering the entire Southern sky South of Declination $+30\degree$ in the frequency range 169$-$231\,MHz. All errors are quoted to $1\sigma$, and we adopt the spectral index convention that $S\propto \nu^{\alpha}$, where $S$ is the measured flux density, $\nu$ is frequency and $\alpha$ is the spectral index.

\section{Data Processing} \label{sec:data}

\subsection{Observations}
The GLEAM observations were performed in a drift-scan observing mode, where all tiles were pointed to the meridian, and the sky drifts through the MWA field of view  \citep[see][]{Wayth2015}. This drift-scan observing mode ensures that the MWA maintains a consistent primary beam throughout a given observing run. Observations were taken during four epochs between 2013 August and 2014 June. Each epoch covered approximately eight hours in Right Ascension, with a $\sim2-4$ hour overlap between epochs. The entire sky South of Declination $+30\degree$ was covered in seven `Declination strips'. The details of each GLEAM observing run are presented by \citet{HurleyWalker2017} and \citet{Lenc2018}.

In completing POGS, we have opted to use the top two GLEAM frequency bands, covering the contiguous frequency range $169-231$~MHz. This frequency range provides a balance between achieving improved resolution in both image space and Faraday space while reducing the impact of depolarisation and retaining sensitivity to large RMs. We note that the MWA beam model is less accurate at these frequencies, and we suffer increased polarisation leakage compared to lower frequencies \citep[e.g.][]{Sutinjo2015,Lenc2016}.

\subsection{Data Reduction}
Calibration and imaging were performed with the RTS \citep{Mitchell2008} using the process detailed by \citet{Lenc2017,Lenc2018} and discussed in \citetalias{Riseley2018}, performed separately on each frequency band. Archival online flagging \citep{Offringa2012} was applied to mitigate radio frequency interference (RFI). Gains were derived for a single calibrator scan on each GLEAM night, selected to maximise signal-to-noise ratio. Images were generated using the RTS, employing Briggs weighting \citep{Briggs1995} using \verb|robust|~$=-1.0$ to reduce sidelobe confusion. We selected baselines in the range $50\lambda$ to $1{\rm{k}}\lambda$ in order to (i) reduce contamination from diffuse Galactic foreground emission and (ii) maintain near-constant resolution for a given Declination strip across the entire frequency band. Image cubes were generated using the native GLEAM channel resolution (40~kHz) for a $20\times20$ deg$^2$ region on a per-snapshot basis, with a 40 arcsecond pixel size to ensure our PSF is oversampled by a factor $\sim5$. Note that the RTS does not perform deconvolution without a-priori knowledge of the source population. While deconvolution is not critical in our case, as the surface density of linearly-polarised sources is sufficiently low to avoid confusion and the majority of polarised sources remain unresolved at the resolution of the Phase I MWA, it can help mitigate leakage due to sidelobes from brighter Stokes $I$ sources in the field.

We then applied corrections for instrumental leakage to correct for inaccuracies in the MWA beam model \citep{Sutinjo2015}, as described in Section 2.1 of \citetalias{Riseley2018}. This correction uses the snapshot nature of GLEAM to reconstruct an empirical `leakage surface' from the apparent Stokes $Q$ and $U$ flux densities as sources drift through the beam \citep[see][for full details]{Lenc2017,Lenc2018}. In a similar manner, a flux scaling correction was derived and subsequently applied by comparing apparent flux densities for sources with entries in the GLEAM catalogue \citep{HurleyWalker2017}. This corrects for position-dependent flux calibration errors that were noted by \citet{HurleyWalker2014,HurleyWalker2017}. 

Following instrumental leakage correction, we applied a correction for ionospheric Faraday rotation using the \verb|RMextract| tool \citep{Mevius2018}, as per Section 2.2 of \citetalias{Riseley2018}. The ionospheric RMs reported by \verb|RMextract| for a given epoch were used to perform a frequency-dependent de-rotation of the Stokes $Q$ and $U$ image cubes on a per-snapshot basis. We note that the expected residual ionospheric RM correction error is of the order of 0.1$-$0.3~rad~m$^{-2}$ for an eight-hour observing run \citep{Sotomayor2013}.

Our leakage- and ionospheric-RM-corrected Stokes $Q$ and $U$ images were then mosaicked using \textsc{swarp} \citep{Bertin2002} to boost our sensitivity (as per Section 2.3 of \citetalias{Riseley2018}), due to the significant overlap between GLEAM snapshots. Finally, the images for each GLEAM frequency band were stacked into one larger cube covering the full frequency range $169 - 231$~MHz.

We then performed RM synthesis \citep[e.g.][]{Burn1966,Brentjens2005} using the CUDA-accelerated Fast Faraday Synthesis (\verb|cuFFS|; \citealt{Sridhar2018}) software. As presented by \citet{Brentjens2005}, the key parameters for RM synthesis -- the RM resolution $(\Delta({\rm{RM}}))$, maximum RM $(|{\rm{RM}}_{\rm{max}}|)$ and maximum scale in RM-space $(\rm{max.~scale})$ -- are defined as 
\begin{subequations}
	\begin{equation}
    	\Delta({\rm{RM}}) = 2 \sqrt{3} / \Delta (\lambda^2) \label{eq:rmparms1}
	\end{equation}
	\begin{equation}
    	{\rm{max.~scale}} = \pi/\lambda_{\rm{min}}^2 \label{eq:rmparms3}
	\end{equation}
	\begin{equation}
	|{\rm{RM}}_{\rm{max}}| = \sqrt{3} / \delta(\lambda^2) \label{eq:rmparms2}
	\end{equation}
\end{subequations}
where $\Delta(\lambda^2)$ is the span in wavelength-squared across the observing bandwidth, $\delta(\lambda^2)$ is the wavelength-squared difference across each observed frequency channel and $\lambda_{\rm{min}}$ is the wavelength of the highest-frequency channel. 

From Equation~\ref{eq:rmparms1}, we have $\Delta({\rm{RM}}) \simeq2.6$~rad~m$^{-2}$, which provides high RM precision. From Equation~\ref{eq:rmparms3} the maximum scale size we can recover is also small, at around 1.9~rad~m$^{-2}$, meaning that all polarised sources will appear point-like in Faraday space \citep[e.g.][]{VanEck2017}. However, from Equation~\ref{eq:rmparms2}, our use of the native 40~kHz GLEAM channel resolution retains sensitivity to very large RMs, up to $\sim1100$~rad m$^{-2}$. For our full POGS sample, we explored Faraday rotations $|{\rm{RM}}| \leq 1000 $ rad m$^{-2}$. We do not anticipate significant bias due to this limit, as typically $|{\rm{RM}}| \ll 200$ rad m$^{-2}$ away from the Galactic plane \citep[e.g.][]{Taylor2009,Schnitzeler2010}. Additionally, no sources in the NVSS RM catalogue have $| {\rm{RM}} | \gtrsim760$ rad m$^{-2}$ \citep{Taylor2009}. Finally, given our long observing wavelength, we also expect complete depolarisation for sources that exhibit variation in RM (i.e. $\sigma_{\rm{RM}}$) at levels much lower than our maximum scale.

We note that, at the time of data processing, \verb|cuFFS| cannot perform deconvolution of the rotation measure spread function (RMSF; analogous to the PSF in aperture synthesis). This technique is better known as `RM-clean' \citep[e.g.][]{Heald2009}. There are two principal areas of improvement that RM-clean can provide: firstly, in recovering structures that are extended in Faraday space (`Faraday-thick'), and secondly, in providing improved precision in the recovered RMs. 

Simulations by \cite{VanEck2018} have demonstrated that low-frequency observations can recover Faraday-thick structures, but that they (i) are heavily depolarised, with $\sim90\%$ of the input flux density unrecovered and (ii) appear as Faraday-thin `skins' on the outer edges of the Faraday-thick structure. Likewise, the dominant contribution to the uncertainty in our RM precision arises not as a result of our RMSF width, but from the precision of the ionospheric RM correction \citep{Sotomayor2013}. Given the relatively shallow depth of our survey and our insensitivity to Faraday-thick structures, and that RM-clean will not provide significant improvement in our RM precision, plus the significant residual leakage in our RM spectra, we opted not to perform RM-clean and instead solely performed RM-synthesis.

\section{Source Identification and Verification}
\subsection{Noise Characterisation}
As discussed in Section 3 of \citetalias{Riseley2018}, we opted to perform source-finding directly on our $P({\rm{RM}})$-cubes. For each epoch, we characterised the noise following the method described by \citet{VanEck2018}. In short, we fitted a Rayleigh distribution to the histogram of $P({\rm{RM}})$ along each pixel, excluding the range $|{\rm{RM}}|\leq 20$ rad m$^{-2}$, as this contained the majority of both residual instrumental leakage as well as residual Galactic foreground contamination.

We measure a typical noise of $\sim1-2$~mJy beam$^{-1}$ RMSF$^{-1}$ near zenith (the Declination $-27\degree$ strip). There is noticeable reduction in sensitivity far from zenith: for example, we measured a typical noise of $\sim4-8$~mJy beam$^{-1}$ RMSF$^{-1}$ for the Declination $-72\degree$ and $+18\degree$ strips, as these pointings lie at $\sim45\degree$ elevation for the MWA.

\subsection{Source Finding}
While the Phase I MWA is limited by confusion noise in Stokes $I$ at the depth reached by GLEAM, the surface density of polarised sources is significantly lower, meaning our image cubes are limited by thermal noise. However, due to a lack of 3D source-finding algorithms in the literature, combined with the non-Gaussian nature of polarisation image noise, we opted to perform `priorised' source-finding, i.e. source finding at the location of known Stokes $I$ sources. In order to strike a balance between probing the faint source population and selecting reliable sources, we took GLEAM sources with a 200\,MHz Stokes $I$ flux density in excess of 90\,mJy. For discussion regarding polarisation source-finding, we refer the reader to \citet{Farnes2018}.

The source-finding process used in POGS (as per Section 3.1 of \citetalias{Riseley2018}) closely follows that described by \citet{VanEck2018}. Regions were extracted centred on the location of known Stokes $I$ sources. The RM spectrum of pixels within the source FWHM (`on-source') was then searched for peaks by identifying local maxima. A peak was considered a `candidate' if it fulfilled three criteria: 
\begin{itemize}
	\item A peak must be in excess of $7{\sigma_{\rm{P}}}$, where ${\sigma_{\rm{P}}}$ is the local noise derived in the previous section. 
	\item A peak must have a flux density greater than the sum of the off-source foreground plus $2{\sigma_{\rm{P}}}$. This off-source foreground is taken as the maximum value of the RM spectrum of pixels below the 1 per cent level of the PSF, centred on the source peak.
	\item A peak must appear outside the instrumental leakage region. This exclusion zone was centred on ${\rm{RM}}=0$~rad~m$^{-2}$, with upper and lower limits defined by the absolute maximum value of the ionospheric RM for that observing epoch. 
\end{itemize}

The enforcement of this third criterion means that we are likely excluding some real polarised sources with low RM. From the NVSS RM catalogue of \cite{Taylor2009}, about 11\% of sources have RMs that would be excluded by this criterion. However, this was a necessary step, given our current inability to fully mitigate instrumental leakage. Future improvements in the MWA beam model may help reduce leakage and allow us to probe this low-RM regime further; however such efforts are beyond the scope of this paper.

Two separate targeted searches were performed. Our primary science goal with POGS was to characterise the low-frequency linearly-polarised extragalactic source population, so the GLEAM Extragalactic Catalogue was used \citep{HurleyWalker2017}. As discussed by those authors (Table 1 of \citealt{HurleyWalker2017}) a handful of sky regions were excluded from the GLEAM catalogue due to data processing issues. The largest region outside the Galactic plane covered 859 square degrees in the region $0{\degree} < {\rm{Declination}} < +30{\degree}$ and $22^{\rm{h}} < {\rm{RA}} < 0^{\rm{h}}$. From visual inspection, we found a handful of polarised sources in this region; a more robust search was performed using the catalogue from the First Alternative Data Release from the TIFR-GMRT Sky Survey \citep[TGSS-ADR1;][]{Intema2017}. A second targeted search was performed across the entire sky, South of Declination $+30\degree$ using the ATNF Pulsar Catalogue \citep{Manchester2005} v1.59 to hunt for known pulsars.

\subsection{Candidate evaluation}
As per Section 3.2 of \citetalias{Riseley2018}, for each candidate source identified by our routine, we fitted a 3D Gaussian (Right Ascension/Declination/RM) using a nine-parameter model, chosen to match the expected form of a source that is unresolved in both image space (as any extended polarised emission from an extragalactic source will tend to rapidly depolarise at our $3^{\prime}-7^{\prime}$ resolution) and Faraday space (as Equations~\ref{eq:rmparms1} and \ref{eq:rmparms3} suggest that any sources detected by the MWA will appear point-like). The nine parameters were:
\begin{itemize}
    \item Broad-band peak polarised intensity $(P)$
    \item Background polarised intensity $(C)$
    \item Image-plane centroids in pixel coordinates $(X,Y)$
    \item Image-plane semi-major $(\sigma_{\rm{maj}})$ and semi-minor $(\sigma_{\rm{min}})$ axes, measured as Gaussian $\sigma$
    \item Image-plane position angle $(PA)$
    \item RM centroid $({\rm{RM}})$
    \item RM width $(\sigma_{{\rm{RM}}})$, measured as Gaussian $\sigma$    
\end{itemize}

Note that the ``background polarised intensity'' is not the same as the foreground discussed in the previous section, but rather the `zero level' of the 3D Gaussian fitted to a candidate. We optimised our model using the \texttt{scipy} `curve-fit' algorithm, employing a Levenberg-Marquardt solver, with initial parameter estimates defined by the initial peak identification. Our candidate list contained a large number of sources that were clearly identified (by visual inspection) as sidelobes of the main instrumental leakage peak, rather than any astrophysical polarised signal. These were frequently poorly-constrained or failed to fit, and were eliminated from our catalogue.

To quantify the uncertainties on our fitted model, as described in Section 3.3 of \citetalias{Riseley2018}, we followed the method described by \citet{VanEck2018}. We established a Monte-Carlo (MC) simulation, performing 1000 realisations of noise, adding the best-fit candidate source model, and re-fitting our nine-parameter model. The standard deviation of the fit results was then used to estimate the uncertainty. Note that the theoretical uncertainty on the measured RM of a source is inversely proportional to the signal-to-noise ratio (SNR) of the detection \citep[e.g.][]{Brentjens2005} according to:
\begin{equation}\label{eq:drm}
    \delta{{\rm{RM}}} = \frac{ \Delta({\rm{RM}}) }{ 2 \times {\rm{SNR}} }
\end{equation}
We found that our MC uncertainties are comparable to the theoretical uncertainty for candidates with high SNR (consistent with previous findings by, e.g. \citealt{George2012}). For candidates with lower SNR $(\lesssim 20)$ the MC uncertainties are typically $30-50$ per cent larger than predicted by Equation~\ref{eq:drm}.

Of all candidates that were successfully fitted, we then rejected any with fitted sizes more than $2\times$ the extent of the PSF or RMSF. Our motivation behind this filtering was twofold: firstly, given the low-frequency nature of our observations, we are only sensitive to Faraday-thin sources (as indicated by Equations \ref{eq:rmparms3} and \ref{eq:rmparms3}). Secondly, due to our moderate resolution with the Phase I MWA, any extended extragalactic polarised sources would likely exhibit RM variance within the PSF (whether due to intrinsic or foreground variation) and rapidly depolarise. As a result of this filtering, plus rejecting fits with poorly-constrained parameters, some $\sim35\%$ of candidates were rejected.

Note that we did not enforce a criterion that the polarised peak must be coincident with the Stokes $I$ peak. Many single sources that appear compact in Stokes $I$ become resolved into multiple components at higher resolution, and it was frequently the case that a polarised source offset from the GLEAM Stokes $I$ peak was in fact associated with one of these components. 

As a final verification step, each fitted candidate was visually inspected in both total intensity and polarisation, with the associated RM spectrum and fitted peak identification. This step was necessary to remove many spurious candidates and chance alignments with sparsely-sampled patches of diffuse Galactic polarised emission. All candidates that conformed to these criteria and passed our tests were considered real, and are henceforth referred to as `sources'. This comprised some $\sim5\%$ of initial identifications.

\subsection{Final measurements}
There is significant overlap between differing epochs of the same Declination strip (some $\sim2-4$h in Right Ascension) as well as overlap between different Declination strip observations of the same hour angle range. As such, a total of 63 sources among our GLEAM-selected population were detected in more than one epoch. For all such sources, the final flux density and RM were determined using the mean of all detections. In all cases of multiple detections, the peak RMs were found to be consistent (to within the $1\sigma$ uncertainties) between epochs. However, we note some variation, up to around $\sim30\%$, in the measured peak polarised flux density between epochs. 

We believe that this variation may be tied to some second-order ionospheric effect that is not yet accurately quantifiable or correctable (priv. comm. Cameron Van Eck \& the LOFAR Magnetism Key Science Project). Investigating the cause of this effect is beyond the scope of this work. Under the assumption that the measured background is largely dominated by noise, we determined final polarised flux density measurements for each source through the quadrature subtraction of the noise from the fitted flux density \citep[for example][]{George2012}. As a final step, we manually compared our nominally extragalactic catalogue with our pulsar catalogue to remove duplicate detections. 

\section{Results}\label{sec:results}
In total, we detect 517 linearly-polarised radio sources: 33 of these are known pulsars, the remaining 484 are assumed to be extragalactic in origin. We thus present two catalogues from this work: POGS ExGal (containing our detected extragalactic radio sources) and POGS PsrCat (containing the known pulsars we have detected). 

\subsection{Extragalactic Radio Sources}
We have attempted to make each catalogue `value-added' by including a number of cross-identifications that the end user may find helpful, and we have also ensured that our catalogue is compliant with the ongoing community effort to standardise reporting of polarisation catalogue data\footnote{\url{https://github.com/Cameron-Van-Eck/RMTable}}. To that end, POGS ExGal contains the following columns:
\begin{itemize}
    \item POGS ID
    \item Right Ascension (J2000) [degrees]
    \item Declination (J2000) [degrees]
    \item Galactic Longitude [degrees]
    \item Galactic Latitude [degrees]
    \item Position uncertainty [degrees]
    \item 200\,MHz Rotation Measure \& uncertainty [rad m$^{-2}$]
    \item Faraday complexity flag ($\ast$)
    \item Faraday complexity identification method ($\ast$)
    \item RM determination method ($\ast$)
    \item Ionospheric RM correction method ($\ast$)
    \item Number of RM components ($\ast$)
    \item 200\,MHz Stokes $I$ flux density \& uncertainty [Jy]
    \item Spectral index \& uncertainty (\emph{see below})
    \item Stokes $I$ reference frequency ($\ast$)
    \item 200\,MHz linear polarisation flux density \& uncertainty [Jy]
    \item Polarisation bias correction method ($\ast$)
    \item Polarised flux density type ($\ast$)
    \item Fractional polarisation \& uncertainty
    \item Linear polarisation reference frequency ($\ast$)
    \item Beam major axis, minor axis, and position angle [degrees]
    \item Beam reference frequency ($\ast$)
    \item Minimum and maximum frequency [Hz] ($\ast$)
    \item Channel width [Hz] ($\ast$)
    \item Number of channels
    \item Channel noise ($\ast$)
    \item Telescope used ($\ast$)
    \item Polarisation catalogue reference ($\ast$)
    \item Stokes $I$ catalogue reference
    \item Stokes $I$ catalogue ID
    \item GLEAM 4Jy Sample ID
    \item Morphological classification (\emph{see below})
    \item Stokes $I$ local rms [Jy beam$^{-1}$]
    \item Linear polarisation local rms [Jy beam$^{-1}$]
    \item NVSS RM catalogue counterpart
    \item 1.4\,GHz NVSS RM \& uncertainty [rad m$^{-2}$]
    \item S-PASS/ATCA RM catalogue counterpart
    \item 2.2\,GHz S-PASS/ATCA RM \& uncertainty [rad m$^{-2}$]
    \item Host Flag (\emph{see below})
    \item Host ID (\emph{see below})
    \item Host Redshift (\emph{see below})
    \item Galactic Foreground RM \& uncertainty [rad m$^{-2}$]
\end{itemize}
where all columns listed above that are marked with $\ast$ have identical values for all sources. These are given in Table~\ref{tab:const_parms}.

\begin{table*}
\centering
\begin{threeparttable}
\caption{List of columns in POGS ExGal and POGS PsrCat that are fixed for all sources. \label{tab:const_parms}}
\footnotesize
\begin{tabular}{|r|l|}
\hline
Column Description & Value \\
\hline
Faraday complexity flag & `N' \\
Faraday complexity identification method & `Inspection' \\
RM determination method & `RM Synthesis - Pol. Int' \\
Ionospheric RM correction method & `RMextract' \\
Number of RM components & 1 \\
Stokes $I$ reference frequency & 200\,MHz \\
Polarisation bias correction method & `2012PASA...29..214G' \\
    & \citep{George2012} \\
Peak or integrated flux density & `Peak' \\
Polarisation reference frequency & 200\,MHz \\
Beam size reference frequency & 200\,MHz \\
Minimum frequency & 169\,MHz \\
Maximum frequency & 231\,MHz \\
Channel width & 40\,kHz \\
Channel noise\tnote{a} & NaN \\
Telescope & `MWA' \\
Catalogue reference\tnote{b} & `2020PASA...37...29R' \\
Data reference\tnote{c} & `2015PASA...32...25W' \\
    & \citep{Wayth2015} \\
\hline
\end{tabular}
\begin{tablenotes}
\item[a]{The rms noise was not measured on a per-channel basis, but from our RM cube, as discussed in the text.}
\item[b]{This is the final bib code of this manuscript.}
\item[c]{This column is intended to provide a link between the catalogue and the paper describing the origin survey.}
\end{tablenotes}
\end{threeparttable}
\end{table*}

For 465 sources in POGS ExGal, we use the spectral index and fitted 200\,MHz Stokes $I$ flux density (with their uncertainties) from the GLEAM Extragalactic Catalogue (respectively `\verb|alpha|' and `\verb|int_flux_fit_200|'; \citealt{HurleyWalker2017}). However, 12 POGS ExGal sources have fitted Stokes $I$ flux densities with significant fractional uncertainty (50\% or greater). Additionally, six sources in POGS ExGal lay in one of the gaps in the GLEAM catalogue (Figure~11 and Table~3 of \citealt{HurleyWalker2017}). These were identified using the 150\,MHz TGSS-ADR1 catalogue as location prior, and thus do not have measured 200\,MHz Stokes $I$ flux densities.

For these 18 sources, we estimated a 200\,MHz flux density using a power-law spectral energy distribution (SED) fit to flux density measurements from the literature. We selected those GLEAM measurements from \cite{HurleyWalker2017} that were reliably constrained, plus data from the following surveys, where available:
\begin{itemize}
    \item 74\,MHz VLA Low-Frequency Sky Survey redux \citep[VLSSr;][]{Lane2014}
    \item 150\,MHz TGSS-ADR1 \citep{Intema2017}
    \item 365\,MHz Texas Radio Survey \citep[TXS;][]{Douglas1996}
    \item 408\,MHz Molonglo Reference Catalogue \citep[MRC;][]{Large1981}
    \item 1.4\,GHz NVSS \citep{Condon1998}
    \item 4.85\,GHz Green Bank 6cm survey \citep[GB6;][]{Gregory1996}
\end{itemize}
The resulting SED fits for the 12 GLEAM sources and six TGSS-ADR1 sources are respectively shown in Appendix~\ref{sec:appendix_a}.

\subsubsection{Host Identification}
Host identification was performed using the same method as for the GLEAM 4-Jy (G4Jy) Sample \citep[see][]{White2018,White2020b,White2020a}. For each source, we created an overlay with GLEAM and higher resolution survey data (specifically the TGSS-ADR1, SUMSS and NVSS where available) superimposed on Widefield Infrared Survey Explorer (\emph{WISE}; \citealt{Wright2010}) mid-infrared images. These were visually inspected, attempting to associate the radio emission to a core galaxy. Once a core was identified, we inspected a \emph{WISE} Band 1 ($3.4\,\upmu$m) cutout at that location and selected the most likely host, where one could be identified. The overwhelming majority of identified hosts had entries in the AllWISE catalogue \citep{Cutri2013}; a handful had clear hosts not catalogued in AllWISE; in these instances, we used entries from the \emph{WISE} catalogue. It is noted in POGS ExGal when this is the case. With the `Host Flag' column, we thus adopt the same formalism as for the G4Jy catalogue, where:
\begin{itemize}
    \item `i': sources which have a clearly identified \emph{WISE}/AllWISE catalogue entry.
    \item `u': sources which do not have a clear \emph{WISE}/AllWISE catalogue entry, either due to the complexity of the Stokes $I$ radio emission or the distribution of nearby \emph{WISE}/AllWISE sources.
    \item `m': sources for which a host galaxy cannot be identified, either due to being too faint to identify in the AllWISE survey, or due to contamination by nearby, bright mid-infrared sources.
\end{itemize}

The \emph{WISE} catalogue position was then cross-referenced with other surveys to determine a redshift, where possible. We used the following catalogues: the Million Quasars catalogue v6.3 \citep[Milliquas;][]{Flesch2015}, the 6dF Galaxy Survey Redshift Catalogue Data Release 3 \citep[6dF;][]{Jones2009}, the Sloan Digital Sky Survey Photometric Catalogue, Release 12 \citep[SDSS DR12;][]{Alam2015} and a number of other specific studies \citep{Simpson1993,Ellison2008,Khabibullina2009}.

\subsubsection{Morphological Classification}
For the morphological classification, we again adopt the same formalism as for the G4Jy Sample. We used the same overlays described above to classify our sources according to the following:
\begin{itemize}
    \item `single': sources which have simple (typically compact) morphologies in higher-resolution data (TGSS-ADR1/SUMSS/NVSS).
    \item `double': sources which have two lobe-like components identified in higher-resolution data (TGSS-ADR1/SUMSS/NVSS) or show elongated structure (suggesting multiple components) but a single catalogued entry in the TGSS-ADR1/SUMSS/NVSS catalogues.
    \item `triple': sources which have two clear lobes in higher-resolution surveys, as well as a clear detection of a core in the same survey.
    \item `complex': sources which do not clearly fit into any of the above categories.
\end{itemize}

Finally, the Galactic foreground RM \& uncertainty are measured from the all-sky Galactic RM reconstruction of \cite{Oppermann2015} at the location of each POGS ExGal source.

\subsection{Known Pulsars}
As with POGS ExGal, we have attempted to add value to POGS PsrCat by including properties from other catalogues that the user may find useful. As such, POGS PsrCat contains the following columns:
\begin{itemize}
    \item POGS ID
    \item Right Ascension (J2000) [degrees]
    \item Declination (J2000) [degrees]
    \item Galactic Longitude [degrees]
    \item Galactic Latitude [degrees]
    \item Position uncertainty [degrees]
    \item 200\,MHz Rotation Measure \& uncertainty [rad m$^{-2}$]
    \item Faraday complexity flag ($\ast$)
    \item Faraday complexity identification method ($\ast$)
    \item RM determination method ($\ast$)
    \item Ionospheric RM correction method ($\ast$)
    \item Number of RM components ($\ast$)
    \item 200\,MHz Stokes $I$ flux density \& uncertainty [Jy]
    \item Spectral index \& uncertainty (\emph{see below})
    \item Stokes $I$ reference frequency ($\ast$)
    \item 200\,MHz linear polarisation flux density \& uncertainty [Jy]
    \item Polarisation bias correction method ($\ast$)
    \item Polarised flux density type ($\ast$)
    \item Fractional polarisation \& uncertainty
    \item Linear polarisation reference frequency ($\ast$)
    \item Beam major axis, minor axis, and position angle [degrees]
    \item Beam reference frequency ($\ast$)
    \item Minimum and maximum frequency [Hz] ($\ast$)
    \item Channel width [Hz] ($\ast$)
    \item Number of channels
    \item Channel noise ($\ast$)
    \item Telescope used ($\ast$)
    \item Polarisation catalogue reference ($\ast$)
    \item Stokes $I$ catalogue reference
    \item Pulsar catalogue ID
    \item Notes
    \item Number of spectral index components (\emph{see below})
    \item Spin Period [s]
    \item Dispersion Measure [pc cm$^{-3}$]
    \item Catalogue reference RM [rad m$^{-3}$]
\end{itemize}
where all columns listed above that are marked with $\ast$ have identical values for all sources. These are given in Table~\ref{tab:const_parms}. The `Notes' column contains additional information, provided as comma-separated-variable (csv) text. This includes whether a pulsar is a known millisecond pulsar (MSP) according to the literature.

A number of columns in POGS PsrCat are sourced from the GLEAM Pulsar Catalogue \citep{Murphy2017}. These are (i) Stokes $I$ flux density measurements at 200\,MHz, (ii) spectral index (plus the associated uncertainties) and (iii) number of spectral index components. Of the 60 known pulsars detected by \citet{Murphy2017}, we detect 21 in polarisation. We also note one pulsar, PSR~J1747$-$4036, that was undetected by \citet{Murphy2017}, but has a compact radio source within 30~arcsec that was detected in the `GLEAM-II: Galactic Plane' reprocessing by \citet{HurleyWalker2019}. This source is GLEAM~J174749$-$403650, and we suggest that it is PSR~J1747$-$4036. There are a further 11 pulsars in POGS PsrCat that do not have 200\,MHz Stokes $I$ flux density measurements, due to their absence from the GLEAM Pulsar Catalogue or GLEAM Galactic Plane Catalogue. We also note that the spectral index fitted by \citet{Murphy2017} was derived using a single-component power-law fit or a two-component broken power-law fit to the available data, and as such does not always span the same frequency range. Where a pulsar has a broken power-law fit, we quote the spectral index value for the `side' of the break on which our observing frequency range lies.

The remainder of this paper is devoted to the discussion of the properties of sources in POGS ExGal (Section~\ref{sec:exgal}) and POGS PsrCat (Section~\ref{sec:psr}). In the appropriate sections of this paper, we present an excerpt from POGS ExGal and the full POGS PsrCat. Unless otherwise stated, all properties are reported at our central frequency of 200\,MHz. All unfilled text entries (e.g. where a POGS source does not have a counterpart in the NVSS RM catalogue) are marked with a `-'; all unfilled numerical entries (e.g. the 1.4\,GHz RM for a POGS source does not have an NVSS RM catalogue counterpart) list a `\verb|NaN|'. However, for display purposes in this manuscript, all such entries list a `-'. Both catalogues will be made available (in the accompanying online material and through Vizier) upon publication of this manuscript.

\section{Extragalactic Radio Sources}\label{sec:exgal}
As a brief overview, POGS ExGal contains 484 sources with linearly-polarised flux densities between 9.9\,mJy and 1.1\,Jy. Our sources have RMs between $-328.07$~rad~m$^{-2}$ and $+279.62$~rad~m$^{-2}$, with a noticeable dearth of sources at $|{\rm{RM}}|\lesssim6$~rad~m$^{-2}$. The mean uncertainty is $0.38$~rad~m$^{-2}$; with a worst-case uncertainty of $10.65$~rad~m$^{-2}$. We find a total of 80 sources in common with the G4Jy Sample. An excerpt from POGS ExGal is presented in Table~\ref{tab:pogs_exgal}.

We detect a total of ten `polarised doubles', by which we mean a pair of physically-associated polarised sources, in our catalogue. These are discussed later in Section~\ref{sec:example_sources} and presented in Figure~\ref{fig:exgal_doubles} and Figure B1 (Appendix~\ref{sec:appendix_b}). We present the RM spectra of the remaining 464 POGS ExGal sources in Appendix~\ref{sec:appendix_d}.

\subsection{Surface Distribution}
The total sky area covered by POGS ExGal is 25,489 square degrees. Hence, our average surface density is 0.019 per square degree, or one polarised extragalactic radio source per 53 square degrees. This represents a 60\% improvement on the source numbers predicted in \citetalias{Riseley2018}, and a factor $\sim2$ improvement compared to predictions by other MWA polarisation studies \citep{Lenc2017}.

\begin{figure*}
\begin{center}
\includegraphics[width=\textwidth]{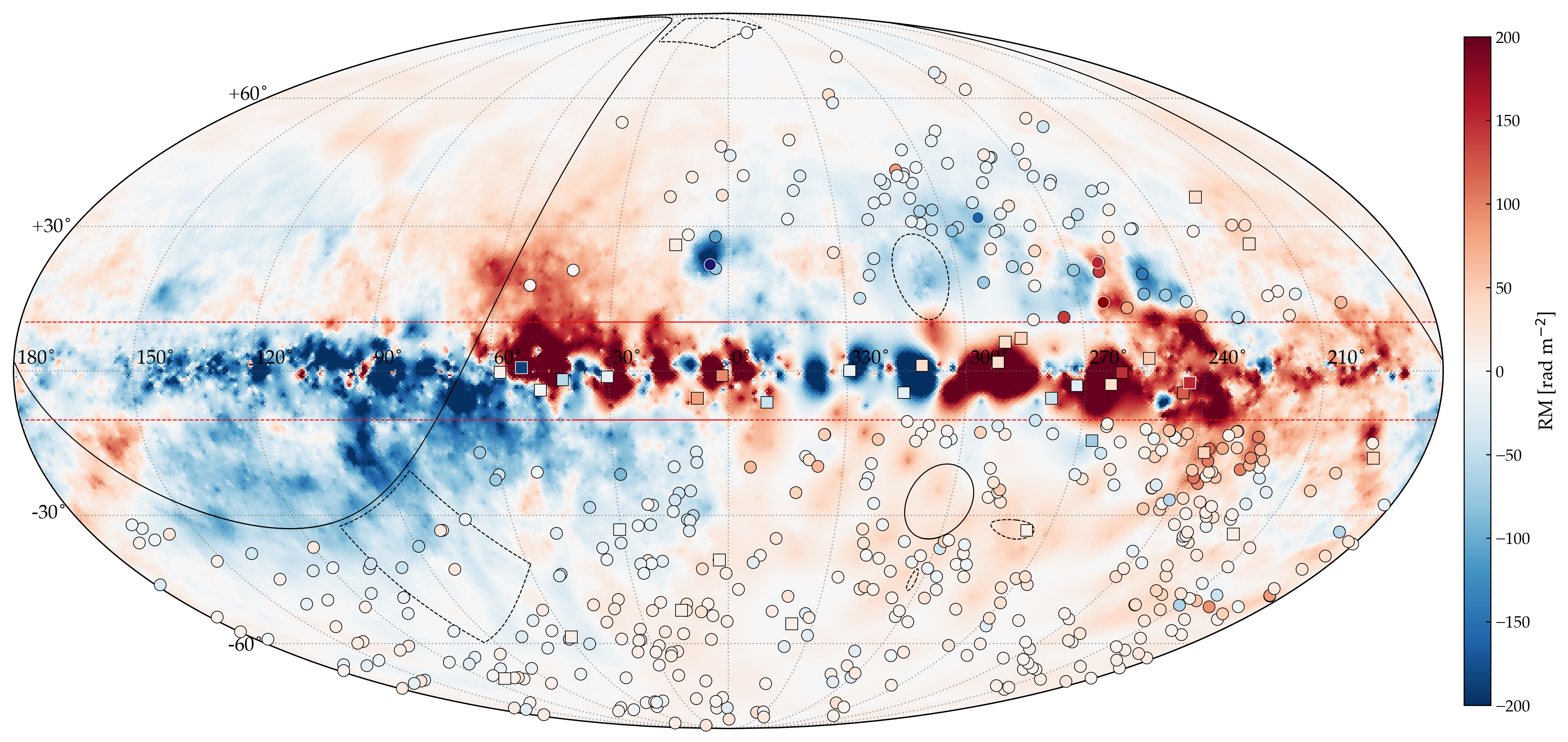} \\
\includegraphics[width=\textwidth]{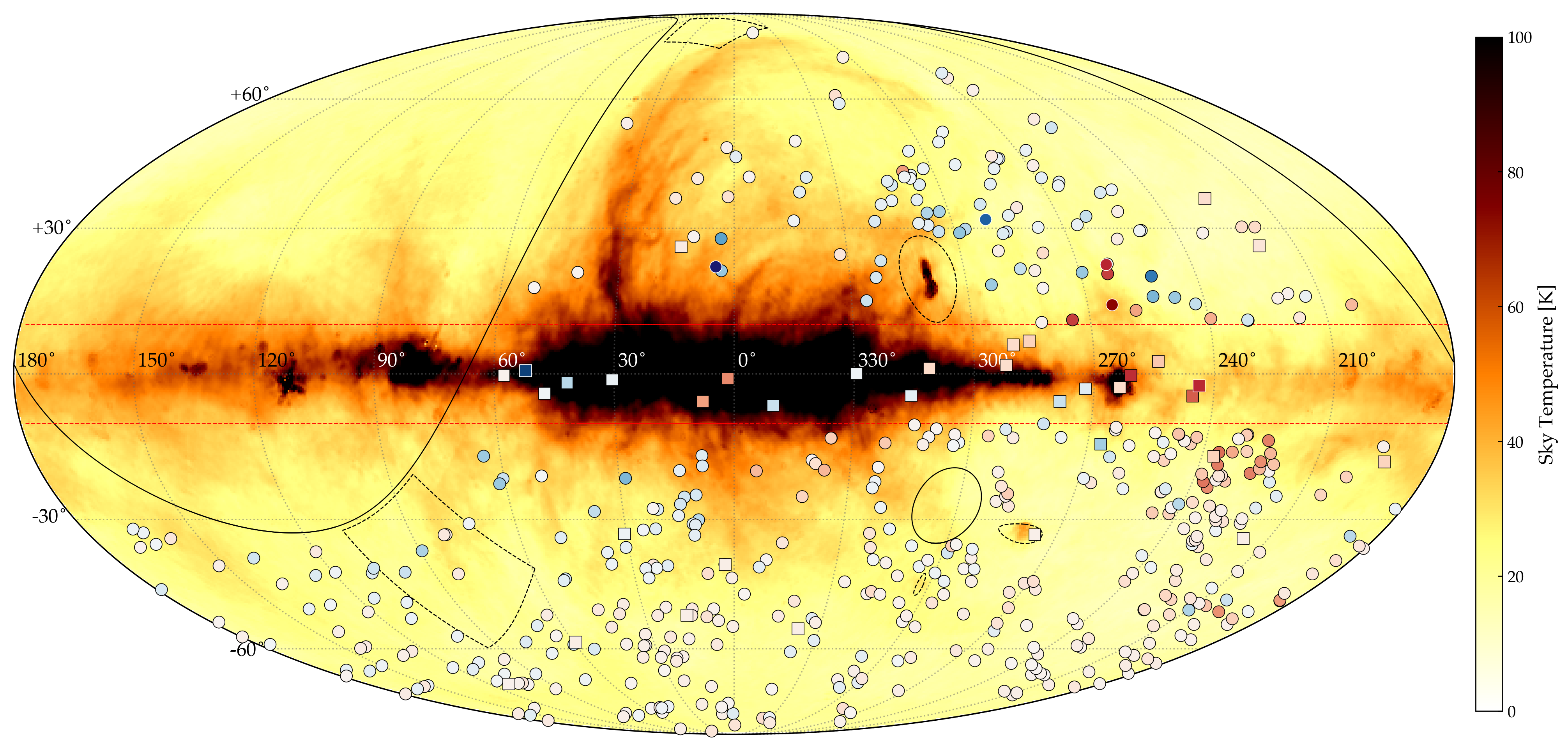}
\caption{Sky surface distribution of sources in POGS ExGal (circles) and POGS PsrCat (squares), shown in Galactic coordinates, colourised according to the sign and magnitude of RM as indicated by the colorbar in the upper panel. The background colourscale shows the Galactic RM from \citet{Oppermann2015} saturating at $|{\rm{RM}}|=200$~rad~m$^{-2}$ (\emph{top panel}) and the 408~MHz Galactic synchrotron emission from \citet{Haslam1982} (\emph{bottom panel}). The solid lines denote the upper and lower Declination limits of our survey coverage ($+30\degree$ and $-82\degree$ respectively); dashed black lines denote gaps in the GLEAM coverage, where source-finding was performed using the TGSS-ADR1 catalogue. Red dashed lines denote the Galactic plane region excluded from the GLEAM Extragalactic Catalogue \citep[$|b| < 10\degree$;][]{HurleyWalker2017}.}\label{fig:skydist}
\end{center}
\end{figure*}

We present the sky distribution for POGS ExGal (circles) and POGS PsrCat (squares) in Figure~\ref{fig:skydist}, overlaid on different maps of the radio sky (the Galactic foreground RM map of \citealt{Oppermann2015} and the all-sky 408\,MHz map of \citealt{Haslam1982}). A number of trends are evident in the distribution of sources. Firstly, there is noticeable decrease in the density of detections toward higher Declinations, with few sources detected at ${\rm{Declination}}\gtrsim+15\degree$. We notice a similar trend toward the south celestial pole, although it is less pronounced. This can be attributed to the reduced sensitivity of the MWA far from zenith, as these pointings were observed at $\sim44\degree$ elevation --- the typical polarised flux density of sources detected in these regions of sky (median $P\sim53\,$mJy) is about a factor two higher than that measured at higher elevations (median $P\sim29\,$mJy). Secondly, away from these low-elevation regions, the distribution of sources appears to be non-uniform. We note a particular clustering of ExGal sources in the region of $(l,b)\sim(230\degree,-20\degree)$, whereas to the Galactic South-West of this region, there is a noticeable dearth of sources around $(l,b)\sim(260\degree,-35\degree)$. Two questions arise: firstly, are these features significant, and if so, what is their physical origin?

\begin{sidewaystable*}

\centering
\small
\caption{Sample rows from POGS ExGal, showing only columns that vary by source. Note that, for display purposes, unfilled entries are marked with a `$-$'. \label{tab:pogs_exgal}}
\begin{tabular}{|crrrrrrrrrr|l}
\hline
POGS ID & RA & Dec & $l$ & $b$ & Pos.Err. & RM & $S_{\rm{200\,MHz}}$ & $\alpha$ & $P_{\rm{200\,MHz}}$ & $\Pi_{\rm{200\,MHz}}$ \\
& $[\degree]$ & $[\degree]$ & $[\degree]$ & $[\degree]$ & $[\degree]$ & [rad m$^{-2}$] & [Jy] & & [Jy] & [\%] \\
\hline
POGSII-EG-001 & $0.11898$ & $-2.86517$ & $94.07524$ & $-62.84601$ & $0.00103$ & $15.08\pm0.14$         & $1.20\pm0.14$ & $-0.98\pm0.02$ & $0.022\pm0.003$         & $1.8\pm0.3$ & ... \\
POGSII-EG-002 & $0.16938$ & $-7.35125$ & $89.28271$ & $-66.84656$ & $0.00073$ & $-13.65\pm0.44$         & $0.58\pm0.12$ & $-0.95\pm0.04$ & $0.023\pm0.004$         & $4.1\pm1.1$ & ... \\
POGSII-EG-003 & $0.55289$ & $-21.88601$ & $55.36275$ & $-77.64491$ & $0.00044$ & $5.91\pm0.04$         & $1.21\pm0.12$ & $-0.82\pm0.02$ & $0.060\pm0.002$         & $5.0\pm0.5$ & ... \\
POGSII-EG-004 & $0.91584$ & $4.01363$ & $100.78891$ & $-56.79011$ & $0.00091$ & $-11.21\pm0.54$         & $0.63\pm0.11$ & $-0.80\pm0.03$ & $0.024\pm0.003$         & $3.7\pm0.8$ & ... \\
POGSII-EG-005 & $1.18757$ & $12.81290$ & $105.61770$ & $-48.48303$ & $0.00064$ & $-14.50\pm0.06$         & $3.84\pm0.31$ & $-0.83\pm0.02$ & $0.131\pm0.006$         & $3.4\pm0.3$ & ... \\
... & ... & ... & ... & ... & ... & ... & ... & ... & ... & ... & ... \\
POGSII-EG-480 & $358.86343$ & $-24.20930$ & $42.85173$ & $-77.05989$ & $0.00066$ & $-10.09\pm0.12$         & $1.28\pm0.11$ & $-0.85\pm0.02$ & $0.019\pm0.003$         & $1.4\pm0.2$ & ... \\
POGSII-EG-481 & $359.17374$ & $-31.82290$ & $8.14227$ & $-77.20659$ & $0.00036$ & $6.90\pm0.08$         & $1.99\pm0.15$ & $-0.61\pm0.02$ & $0.062\pm0.003$         & $3.1\pm0.3$ & ... \\
POGSII-EG-482 & $359.61725$ & $-23.04741$ & $48.82280$ & $-77.34032$ & $0.00090$ & $21.95\pm0.19$         & $0.52\pm0.08$ & $-0.66\pm0.03$ & $0.015\pm0.003$         & $2.8\pm0.7$ & ... \\
POGSII-EG-483 & $359.76191$ & $-23.27486$ & $48.06751$ & $-77.54639$ & $0.00089$ & $-10.29\pm0.35$         & $0.87\pm0.09$ & $-0.79\pm0.02$ & $0.015\pm0.003$         & $1.7\pm0.4$ & ... \\
POGSII-EG-484 & $359.94218$ & $20.58771$ & $106.91553$ & $-40.67160$ & $0.00081$ & $-41.29\pm7.79$         & $1.20\pm0.10$ & $-0.70\pm0.05$ & $0.034\pm0.023$         & $2.9\pm2.0$ & ... \\
\hline
\end{tabular}
\bigskip

\begin{tabular}{r|rrrcccrr|cr|cr|l}
\hline
 & & & & & & & & & \multicolumn{2}{c|}{NVSS RM catalgoue} &  \\
 & Bmaj & Bmin & BPA & DataID & G4Jy ID &  Morph. & $\sigma_I$ & $\sigma_P$ & ID & RM &  \\
 & $[\degree]$ & $[\degree]$ & $[\degree]$ & & & & [Jy] & [Jy] & & [rad m$^{-2}$] & \\
\hline
... & 0.06780 & 0.05572 & 183.3 & GLEAM J000031-025141 & - & S &         0.012 & 0.003 & - &$-$ & ... \\
... & 0.06366 & 0.05339 & 181.3 & GLEAM J000038-072143 & - & D &         0.011 & 0.003 & - &$-$ & ... \\
... & 0.04849 & 0.04815 & 299.2 & GLEAM J000212-215307 & - & S &         0.009 & 0.003 & 000211-215309 & $6.0\pm4.9$ & ... \\
... & 0.06780 & 0.05572 & 183.3 & GLEAM J000340+040047 & - & C &         0.013 & 0.003 & 000340+040102 & $-9.7\pm10.3$ & ... \\
... & 0.09536 & 0.08242 & 262.7 & GLEAM J000441+124907 & G4Jy 7A & D &         0.023 & 0.005 & - & $-$ & ... \\
... & ... & ... & ... & ... & ... & ... & ... & ... & ... & ... &  ... \\
... & 0.04849 & 0.04815 & 299.2 & GLEAM J235527-241227 & - & S &         0.007 & 0.002 & 235527-241226 & $3.3\pm6.2$ & ... \\
... & 0.05250 & 0.04697 & 358.2 & GLEAM J235641-314919 & - & S &         0.007 & 0.003 & 235640-314923 & $14.7\pm1.4$ & ... \\
... & 0.04849 & 0.04815 & 299.2 & GLEAM J235829-230225 & - & D &         0.007 & 0.002 & 235829-230225 & $18.8\pm11.0$ & ... \\
... & 0.04849 & 0.04815 & 299.2 & GLEAM J235900-231629 & - & S &         0.006 & 0.002 & 235900-231630 & $7.3\pm13.5$ & ... \\
... & 0.09536 & 0.08242 & 262.7 & TGSSADR J235943.4+203604 & - & S &         $-$ & 0.005 & 235946+203614 & $-30.9\pm1.7$ & ... \\
\hline
\end{tabular}
\bigskip

\begin{tabular}{r|crc|crcc|r|l|}
\hline
& \multicolumn{3}{|c|}{S-PASS/ATCA RM catalgoue} & \multicolumn{4}{c|}{Host} & \\
 & ID & RM & N(RM) & Flag & ID & Cat & Redshift & ${\rm{RM}}_{\rm{Gal}}$ \\
 & & [rad m$^{-2}$] & & & & & & [rad m$^{-2}$] \\
\hline
... & - & $-$ & 1 &         i & J000031.60-025150.7 & AllWISE & $-$ & $-2.3\pm5.0$  \\
... & - & $-$ & 1 &         i & J000039.82-072129.5 & AllWISE & $-$ & $3.9\pm4.1$  \\
... & PKSB2359-221 & $3.4\pm0.5$ & 1 &         i & J000211.98-215310.0 & AllWISE & $-$ & $5.4\pm2.9$  \\
... & - & $-$ & 1 &         i & J000339.94+040046.4 & AllWISE & $-$ & $-5.0\pm4.5$  \\
... & - & $-$ & 1 &         i & J000450.26+124839.6 & AllWISE & 0.143 & $-14.0\pm4.4$  \\
... & ... & ... & ... & ... & ... & ... & ... & ... \\
... & - & $-$ & 1 &         i & J235527.62-241226.2 & AllWISE & $-$ & $5.1\pm2.9$  \\
... & PKSB2354-321 & $3.5\pm0.5$ & 3 &         i & J235640.81-314923.8 & AllWISE & $-$ & $8.2\pm2.9$  \\
... & - & $-$ & 1 &         i & J235829.31-230223.8 & AllWISE & $-$ & $9.0\pm3.2$  \\
... & - & $-$ & 1 &         i & J235900.77-231630.5 & AllWISE & $-$ & $9.0\pm3.2$  \\
... & - & $-$ & 1 &         i & J235944.51+203606.6 & AllWISE & 0.565 & $-30.8\pm5.0$  \\
\hline
\end{tabular}

\end{sidewaystable*}

\begin{figure*}
\begin{center}
\includegraphics[width=\textwidth]{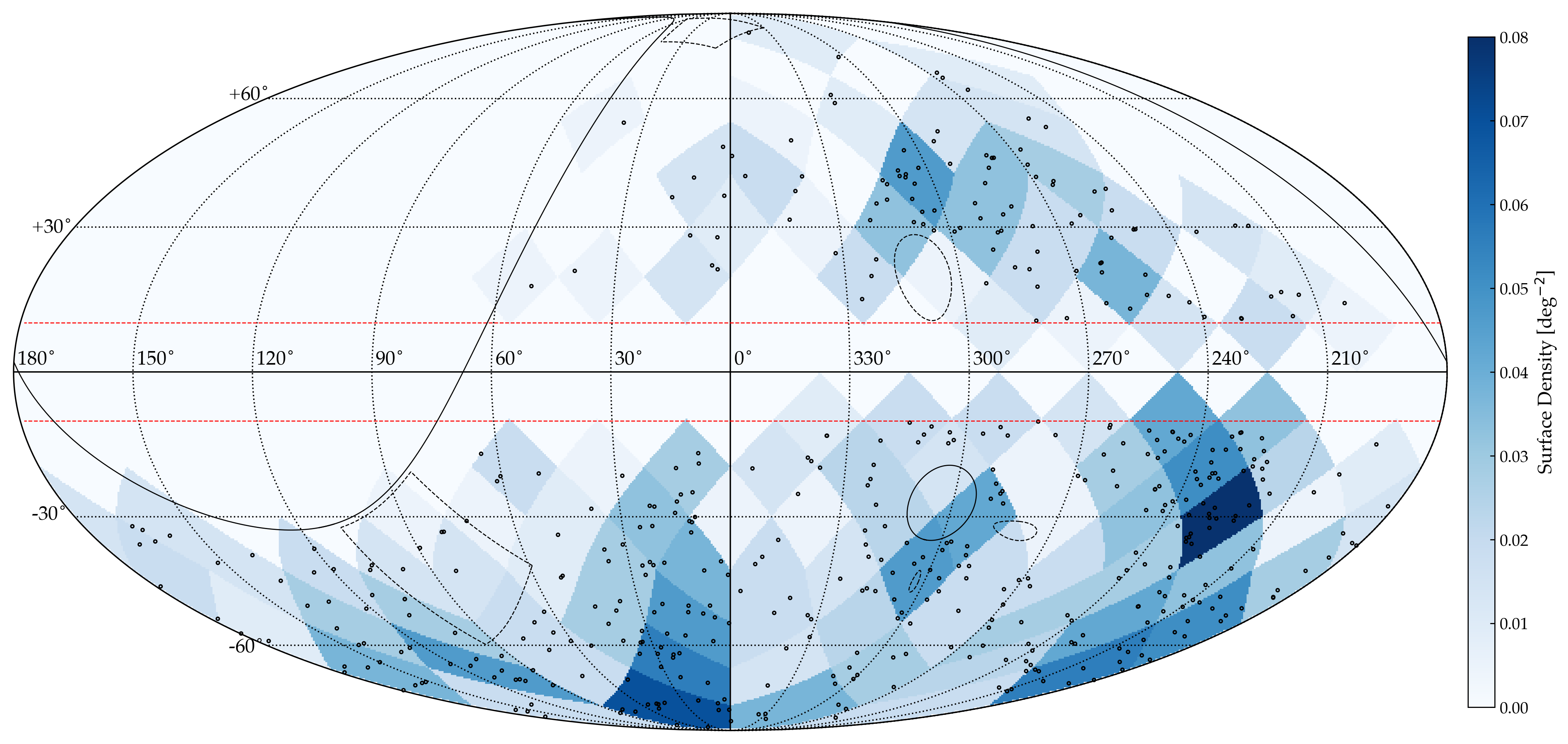}
\includegraphics[width=\textwidth]{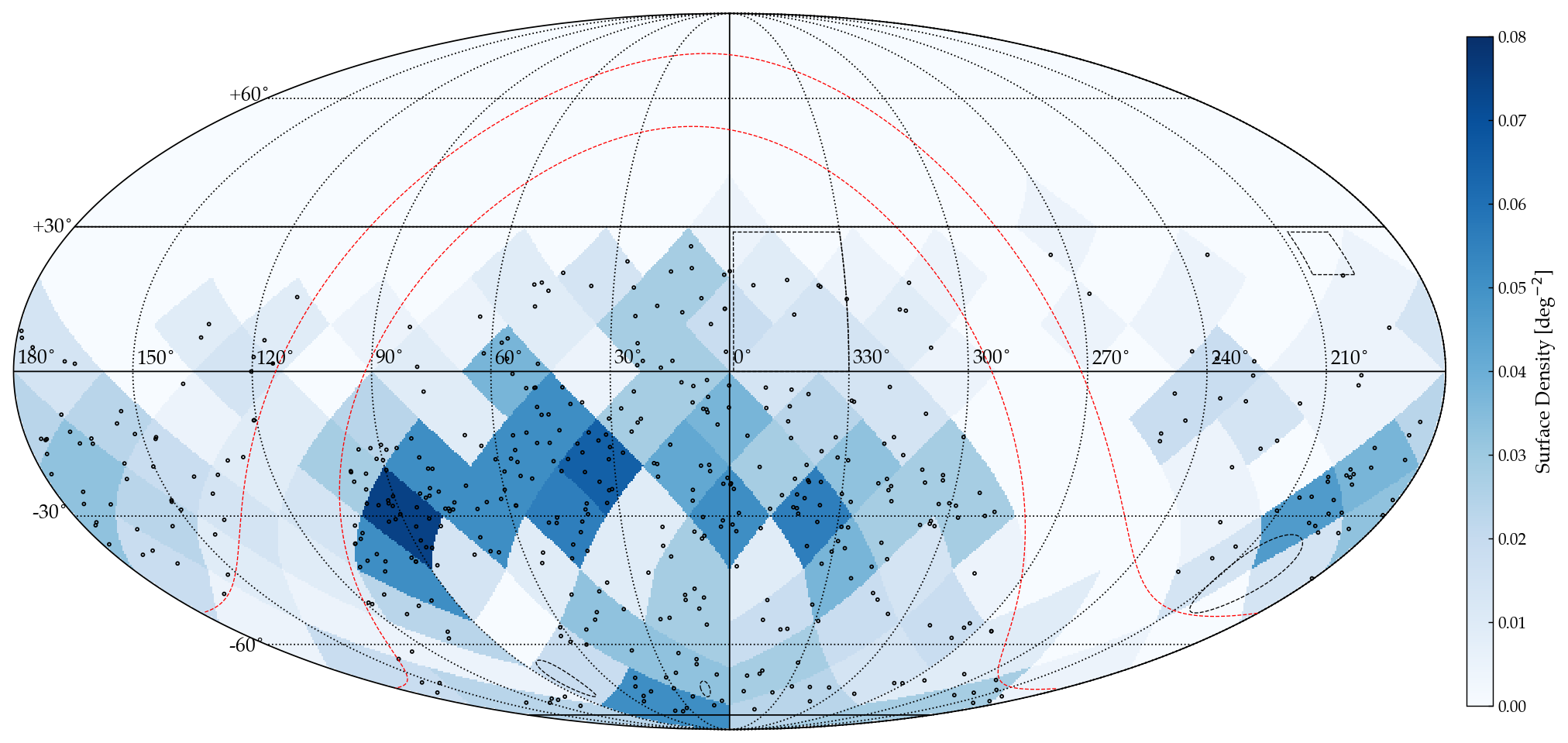}
\cprotect\caption{Surface density of POGS ExGal sources (black circles) derived using a HEALPix Mollweide projection in Galactic coordinates (\emph{top panel}) and Equatorial J2000 coordinates (\emph{bottom panel}) with \verb|NSIDE|~$=4$. Dashed and solid curves denote exclusion regions as per Figure~\ref{fig:skydist}.} 
\label{fig:healpix}
\end{center}
\end{figure*}

\subsubsection{Clustering Significance}\label{sec:clustering}
To quantify the significance of these over- and under-density regions, we divided the sky into a set of tiles using Hierarchical Equal Area isoLatitude Pixelization \citep[HEALPix;][]{HEALPix} with \verb|NSIDE|$=4$. This yields 192 tiles of approximately 214 square degrees covering the entire sky; the surface density is then simply given by the number of sources within a given tile. The surface density of sources is shown in Figure~\ref{fig:healpix}.

From Figure~\ref{fig:healpix}, the surface density appears to vary substantially from tile-to-tile, supporting the suggestion that our sources are not uniformly distributed across the sky. Note that, as suggested by Figure~\ref{fig:healpix}, this appears to be irrespective of coordinate projection, suggesting the distribution is not a related to decreases in instrumental sensitivity away from zenith. From inspection of our rms noise maps, we can also discount the idea that this clustering is simply the result of local noise variations.

We then estimated the median surface density exclusively using those tiles whose central coordinates lay outside the exclusion zones shown in Figures~\ref{fig:skydist} and \ref{fig:healpix}, leading us to discard those tiles at ${\rm{Declination}}\geq+30\degree$ and $|b|\leq10\degree$. We find the resulting median density is 0.014~deg$^{-2}$, with a standard deviation of 0.016~deg$^{-2}$. The apparent under-dense region around $(l,b)\sim(260\degree,-35\degree)$ has a typical density of 0.0047~deg$^{-2}$ which, while much lower than the median, is not significant. This is also visible in Figure~\ref{fig:healpix}, where many tiles exhibit a similar density. 

The `cluster' of sources visually identified in Figure~\ref{fig:skydist} around $(l,b)\sim(230\degree,-20\degree)$ lies within a tile where the source density is 0.05~deg$^{-2}$. Some $5.4\%$ of tiles have a surface density equal to or greater than this, so we do not consider this clustering particularly significant. Only two tiles (out of 112 that do not lie within the `exclusion zones' defined in Figure~\ref{fig:healpix}) show a surface density in excess of $3\sigma$. These are the tile centred on $(l,b) = (230\degree,-30\degree)$, where the surface density is 0.079~deg$^{-2}$ (a $4\sigma$ outlier) and the tile centred on $(l,b) = (45\degree, -78\degree)$, which has a surface density of 0.070~deg$^{-2}$ $(\sim3.5\sigma)$. However, this is still only $\sim2\%$ of tiles in Figure~\ref{fig:healpix}, so we cannot conclusively discount that this is simply random chance.

\subsubsection{Large-RM Sources}
A total of 14 sources in our catalogue have large RM values, defined as $|{\rm{RM}}| > 100$~rad~m$^{-2}$. We summarise the Galactic coordinates, RMs, and associations of these sources in Table~\ref{tab:largerm}. Two of these, GLEAM~J161719$-$100227 and GLEAM~J163927$-$124141 lie along the LOS through the nearby H\,\textsc{i}\textsc{i} region, Sharpless 2-27. Sources along the LOS through this H\,\textsc{i}\textsc{i} region are known to have significant RM enhancement \citep[e.g.][]{Harvey-Smith2011}.

\begin{table*}
\centering
\caption{Population of POGS ExGal sources with large absolute RM values, defined as $|{\rm{RM}}| > 100$ rad m$^{-2}$. \label{tab:largerm}}
\footnotesize
\begin{tabular}{ccrrrp{3cm}}
\hline
POGS ID & Source Name & $l$ & $b$ & RM & Association \\
 & & [deg] & [deg] & [rad m$^{-2}$] & \\
\hline
POGSII-EG-126 & GLEAM~J031522$-$031643 & 184.53202 & $-48.1562$  & $+110.89 \pm 0.11$ & Arc B of Orion-Eridanus Superbubble \\
POGSII-EG-200 & GLEAM~J055905$-$201306 & 225.90924 & $-20.3022$  & $+100.65 \pm 0.85$ & ... \\
POGSII-EG-207 & GLEAM~J060840$-$304130 & 237.21561 & $-22.0079$  & $+106.69 \pm 0.28$ & ... \\
POGSII-EG-214 & GLEAM~J062028$-$274020 & 235.18013 & $-18.5343$  & $+104.66 \pm 0.10$ & ... \\
POGSII-EG-216 & GLEAM~J062213$-$155817 & 224.14857 & $-13.5471$  & $+100.14 \pm 0.13$ & ... \\
POGSII-EG-219 & GLEAM~J063228$-$272109 & 235.91364 & $-15.9525$  & $+111.30 \pm 0.07$ & ... \\
POGSII-EG-263 & GLEAM~J092317$-$213744 & 251.70999 & $+19.9326$  & $-140.98 \pm 0.05$ & H$\alpha$ arc of Gum Nebula \\
POGSII-EG-264 & GLEAM~J094056$-$335914 & 263.78321 & $+14.0358$  & $+279.62 \pm 0.21$ & H$\alpha$ arc of Gum Nebula \\
POGSII-EG-268 & GLEAM~J095750$-$283808 & 262.86672 & $+20.4215$  & $+137.60 \pm 0.16$ & H$\alpha$ arc of Gum Nebula \\
POGSII-EG-271 & GLEAM~J100206$-$265606 & 262.45800 & $+22.3381$  & $+153.02 \pm 0.07$ & H$\alpha$ arc of Gum Nebula \\
POGSII-EG-273 & GLEAM~J101236$-$425901 & 274.51326 & $+10.9617$  & $+138.56 \pm 0.19$ & H$\alpha$ arc of Gum Nebula \\
POGSII-EG-308 & GLEAM~J120238$-$294841 & 290.47462 & $+31.8878$  & $-164.42 \pm 0.58$ & ... \\
POGSII-EG-359 & GLEAM~J161719$-$100227 & 3.49433   & $+27.8050$  & $-107.00 \pm 0.10$ & Sh2-27 \\
POGSII-EG-363 & GLEAM~J163927$-$124141 & 4.78538   & $+21.8817$  & $-328.07 \pm 0.36$ & Sh2-27 \\
\hline
\end{tabular}
\end{table*}

We also detect a single high-|RM| source, GLEAM~J031522$-$031643, which lies coincident with `Arc B' of the Orion-Eridanus Superbubble \citep[e.g.][]{Ochsendorf2015,Soler2018}. The high-|RM| source GLEAM~J120238$-$294841 has no apparent association with any Galactic foreground structure in total intensity, H$\alpha$ or Galactic linear polarisation, from e.g. the S-band Polarisation All Sky Survey \citep[S-PASS;][]{Carretti2019}.

\begin{figure*}
\centering
\includegraphics[width=0.8\textwidth]{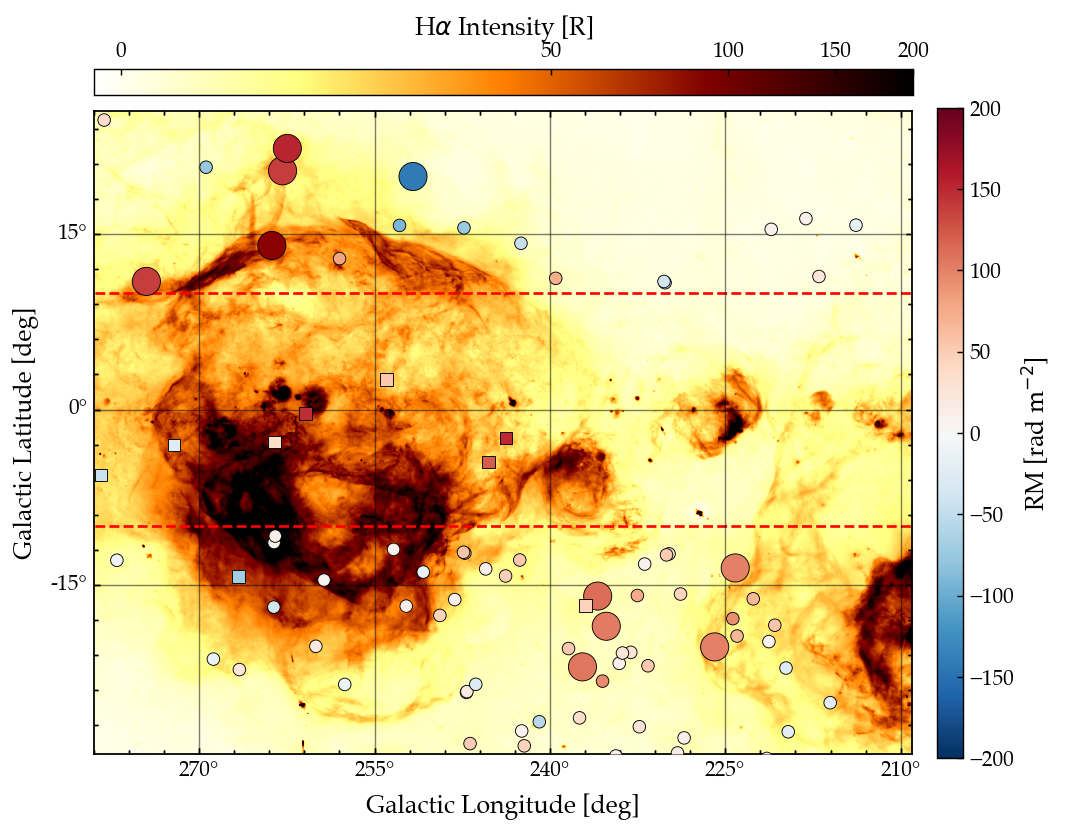}
\caption{Close-up image of the Gum Nebula region. The background image shows H$\alpha$ emission \citep{Finkbeiner2003} on an arcsinh stretch. Circular (square) markers denote POGS ExGal (PsrCat) sources, colourised according to |RM| as indicated by the colour bar. POGS ExGal sources with large absolute RMs, defined as $|{\rm{RM}}| > 100$\,rad~m$^{-2}$, are indicated by the larger markers. Red dashed lines denote the Galactic plane exclusion zone omitted from the GLEAM catalogue.}
\label{fig:gum_nebula}
\end{figure*}

Five sources in our large-RM sample (GLEAM~J092317$-$213744, GLEAM~J094056$-$335914, GLEAM~J095750$-$283808, GLEAM~J100206$-$265606 and GLEAM~J101236$-$425901) lie along multiple LOS through the northern H$\alpha$ arc of the Gum Nebula \citep[e.g.][]{Purcell2015}. We show a close-up of this region in Figure~\ref{fig:gum_nebula}. Note that one of these sources was detected in \citetalias{Riseley2018}. A large magnetic bubble was identified in this region by \citet{Vallee1983} who detected enhanced RMs (up to $|{\rm{RM}}| \sim 200$~rad~m$^{-2}$) within $\sim20\degree$ of the Gum Nebula. The typical absolute RM of these five sources is 160~rad~m$^{-2}$, suggesting that we detect the same feature, although our sample is limited by its proximity to the Galactic plane. 

A further five sources in our large-|RM| sample (GLEAM~J055905$-$201306, GLEAM~J060840$-$304130, GLEAM~J062028$-$274020, GLEAM~J062213$-$155817 and GLEAM~J063228$-$272109) are members of the source `cluster' around $(l,b)\sim(230\degree,-20\degree)$, discussed in Section~\ref{sec:clustering}. The LOS to these sources passes between the Gum Nebula and Barnard's Loop. We have shown that the density of sources in this region is not significantly enhanced compared to the average sky distribution; neither is the mean |RM| significantly higher in this region, as might be expected if there were a magnetic shell similar to that detected to the North of the Gum Nebula.

\subsection{Notes on Individual Sources}\label{sec:example_sources}
\begin{figure*}
\centering
\includegraphics[width=0.8\textwidth]{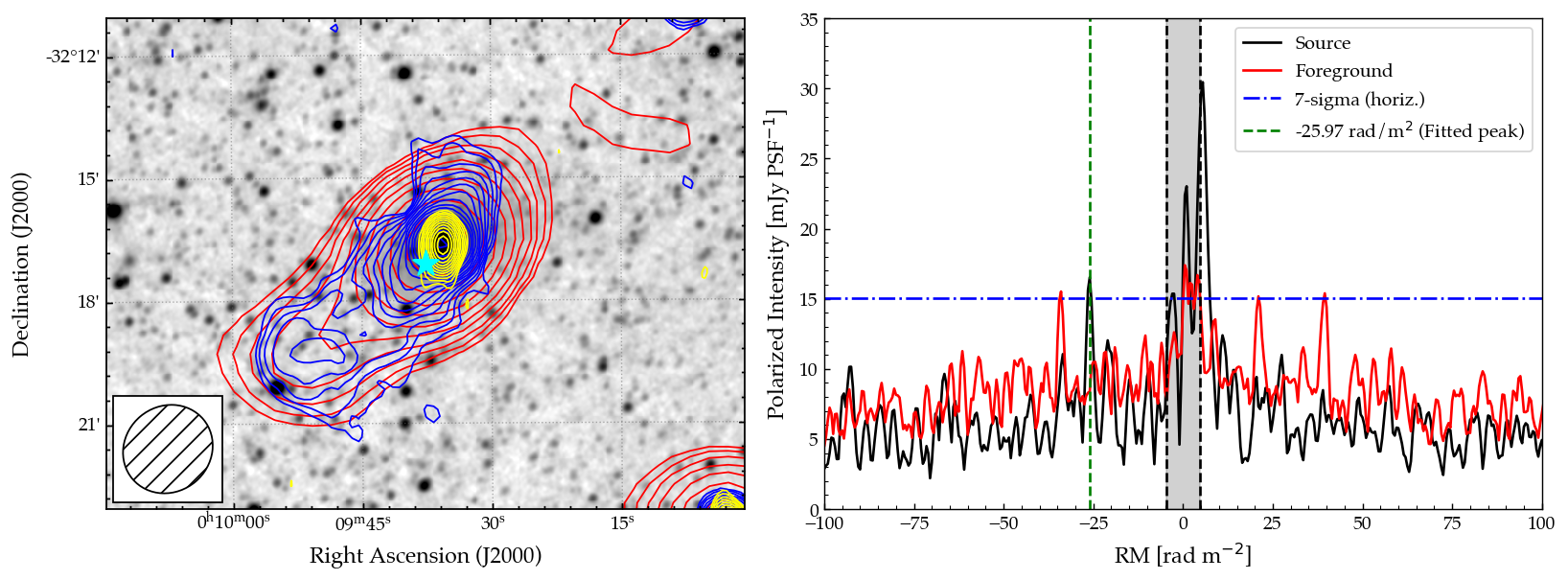}
\includegraphics[width=0.8\textwidth]{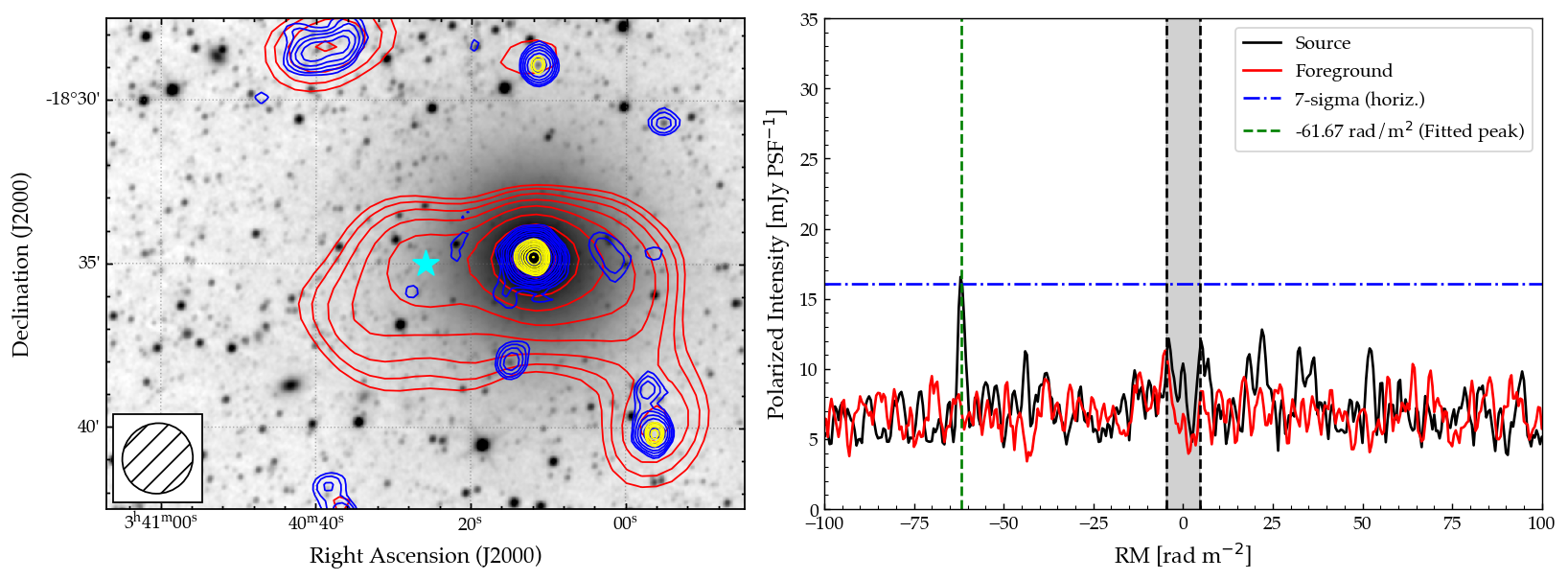}
\includegraphics[width=0.8\textwidth]{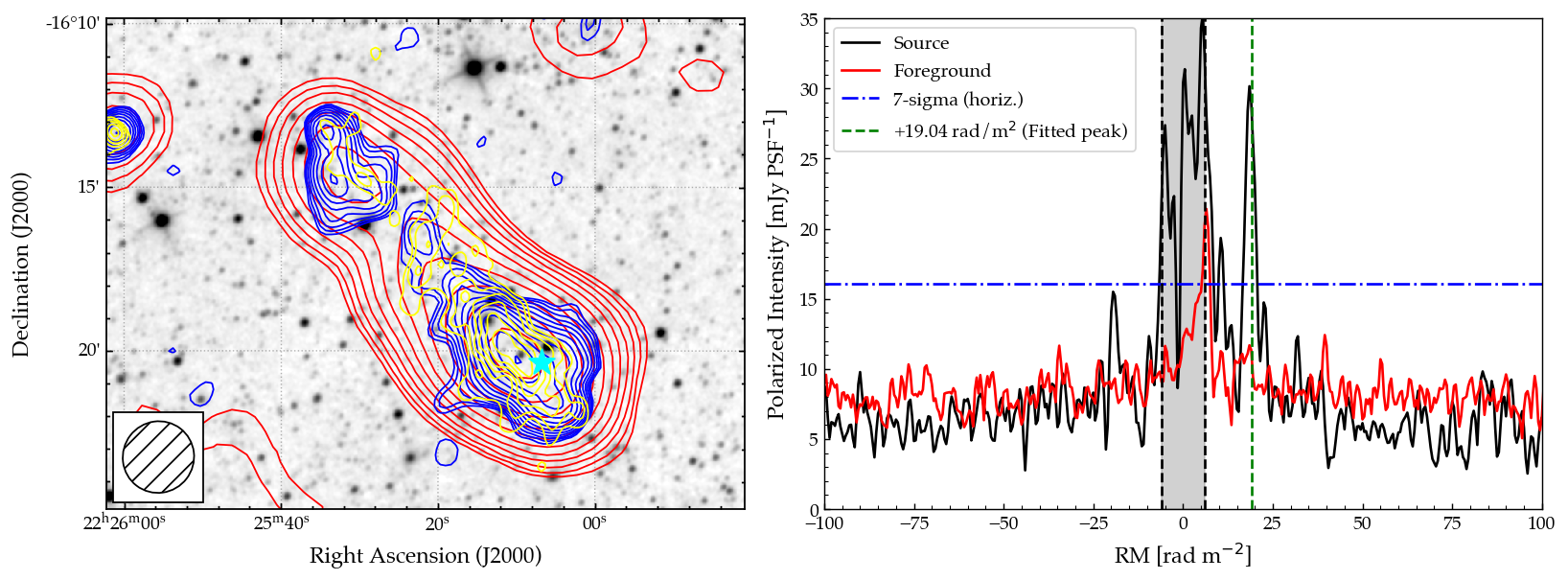}
\includegraphics[width=0.8\textwidth]{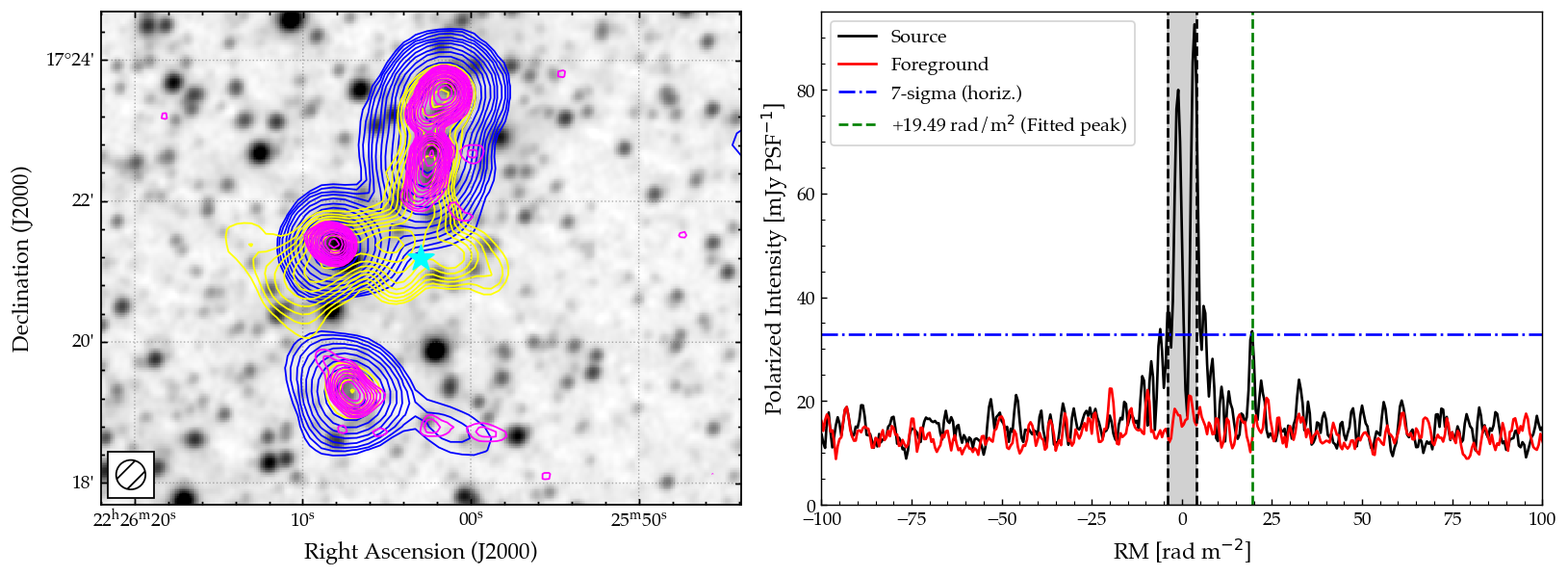}
\caption{Examples of POGS ExGal sources that have complex and/or extended Stokes $I$ continuum morphologies. From top to bottom, sources are GLEAM~J000936$-$321640, GLEAM~J034026$-$183545, GLEAM~J222510$-$162001 and  TGSSADR~J222603.3$+$172208. \emph{Left panels} show \emph{WISE} W1 ($3.4\,\upmu$m) infrared surface brightness in grayscale, with total intensity contours from GLEAM 200\,MHz (red), the TGSS-ADR1 150\,MHz (yellow), and the NVSS 1.4\,GHz (blue). In the lower panel, archival C-configuration VLA data at 1.4\,GHz are overlaid in magenta. Note that this source does not have GLEAM continuum contours as it lies within one of the `gaps' in the GLEAM survey coverage. Cyan stars denote the coordinates of the polarised peak. The resolution of the survey used for the source search (i.e. GLEAM for the first three panels, TGSS-ADR1 for the lower panel) is shown as the hatched ellipse in the lower-left corner. \emph{Right panels} show the source RM spectrum along the LOS through the cyan star (black) plus the foreground RM spectrum (red) as well as the instrumental leakage avoidance zone (shaded gray region). Green dashed line denotes the fitted RM; blue dot-dashed line denotes the $7\sigma$ level.}
\label{fig:exgal_examples}
\end{figure*}

%This is really an excuse to show some of the more fun sources in our sample.
We detect polarised emission from regions of a number of sources that exhibit extended and/or complex Stokes $I$ continuum morphologies. Examples of these are given in Figure~\ref{fig:exgal_examples}, where we show cutouts around the location of polarised emission as well as the RM spectrum along the line of sight through the polarised peak. Note that none of the sources in Figure~\ref{fig:exgal_examples} were detected in polarisation by either \citet{Taylor2009} or \citet{Schnitzeler2019}. 

From the top panel of Figure~\ref{fig:exgal_examples}, the source spectrum of GLEAM~J000936$-$321640 shows a single significant peak outside the leakage zone-of-avoidance, but three peaks in the off-source foreground spectrum that are at the $7\sigma$ level. These foreground peaks are examples of contamination by residual Galactic foreground emission: when inspecting the RM cube, the emission corresponding to these peaks appears extended in the image plane, whereas the emission associated with this source is compact in the image plane. 

\begin{figure*}
\centering
\includegraphics[width=\textwidth]{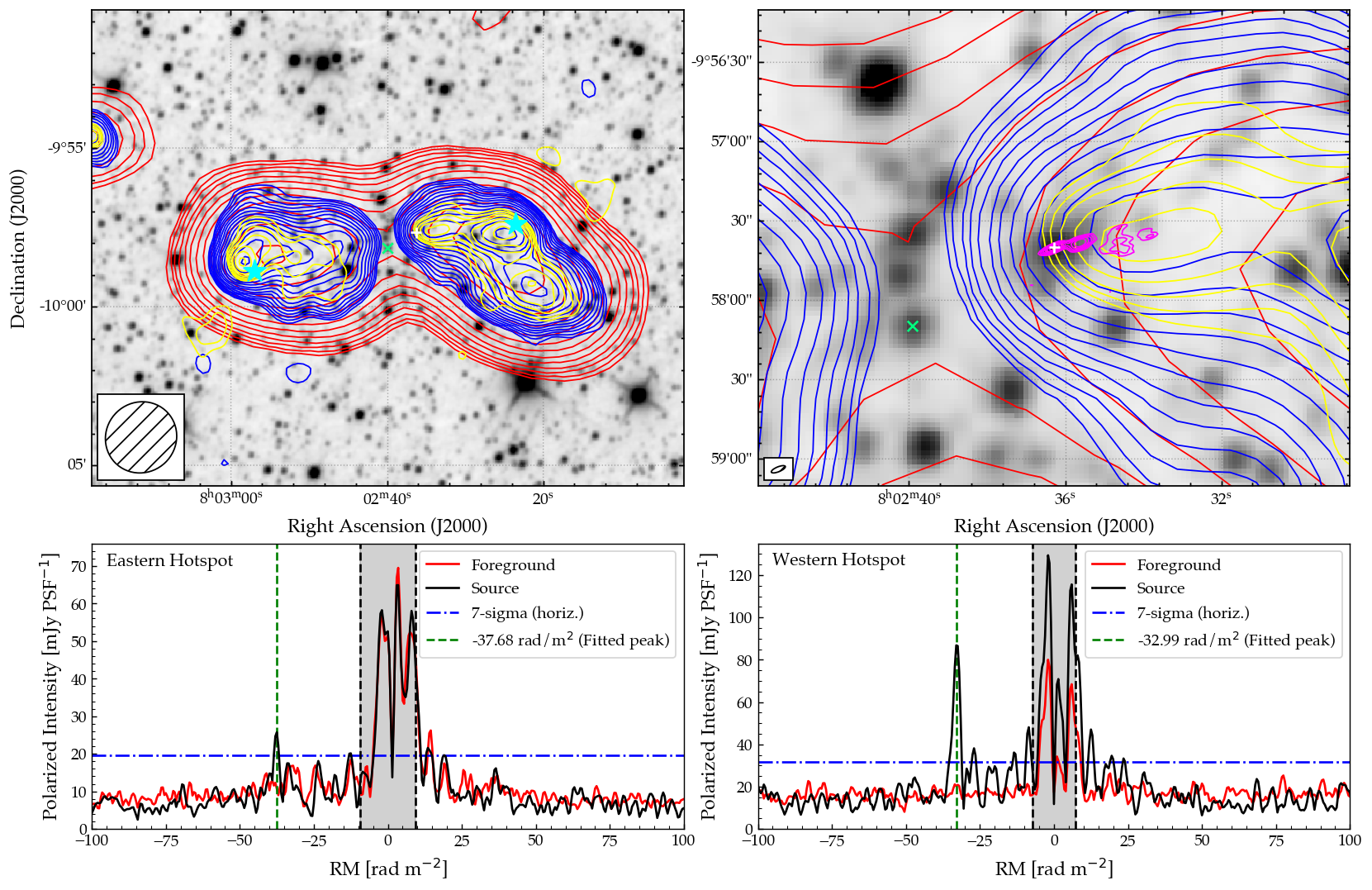}
\includegraphics[width=\textwidth]{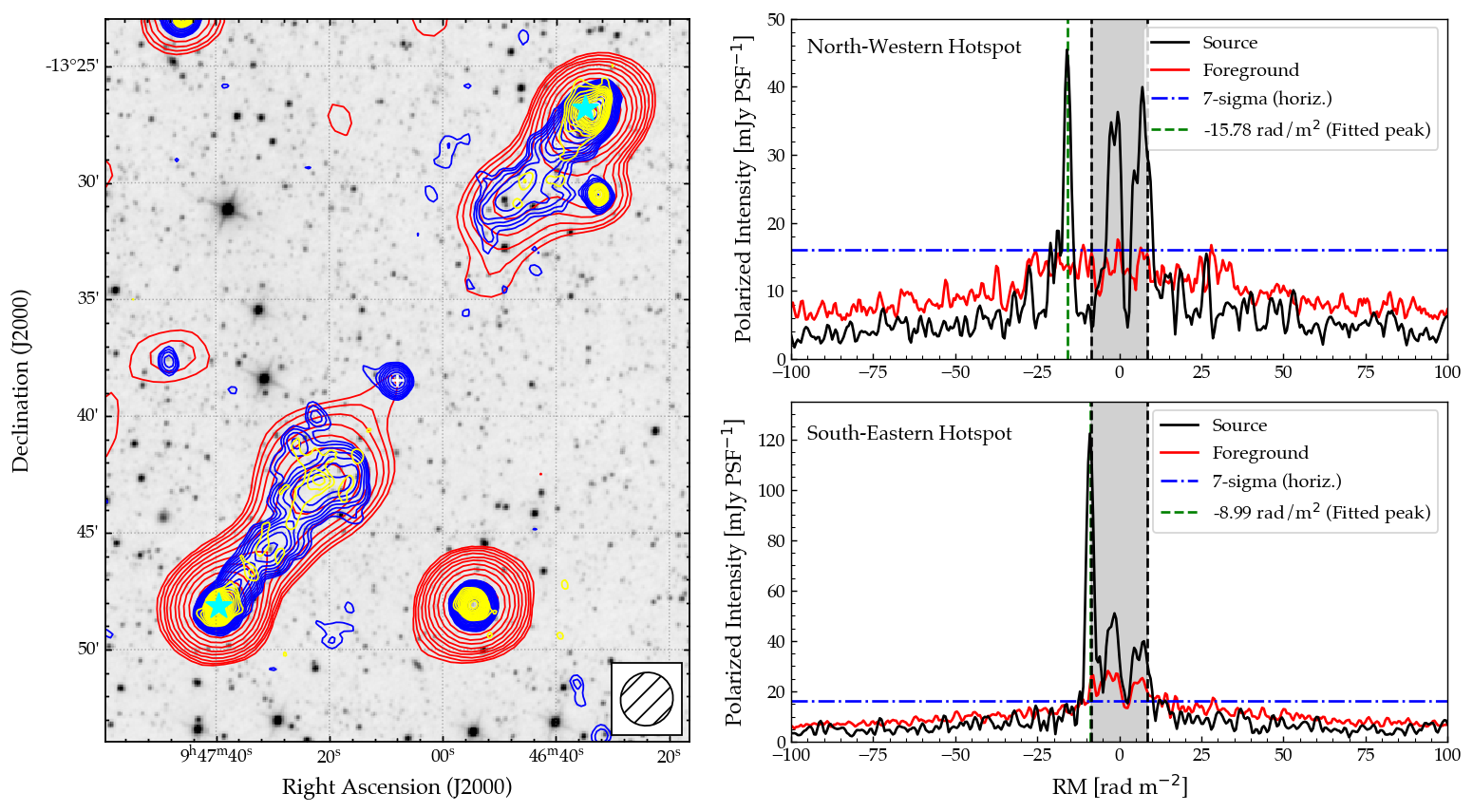}
\cprotect\caption{Examples of `polarised doubles', where two POGS ExGal sources are associated with a single, extended radio galaxy. Sources are POGSII-EG-250 \& POGSII-EG-251 (PKS~B0800$-$09, G4Jy~680; \emph{top}) and  POGSII-EG-265 \& POGSII-EG-266 (`J0947$-$1338'; \emph{bottom}). The grayscale is \emph{WISE} W1 ($3.4\,\upmu$m) infrared surface brightness, and contours are as per Figure~\ref{fig:exgal_examples}. The top-right panel shows a close-up of the core of PKS~B0800$-$09, with archival B/C configuration VLA data at 4.89\,GHz in magenta. These data were used to select the correct host galaxy for this source. Placement of the RM spectrum subplots denotes which LOS in the postage stamp they are shown along. Hatched ellipses denote the resolution of the GLEAM 200\,MHz continuum image.}
\label{fig:exgal_doubles}
\end{figure*}

The lower panel of Figure~\ref{fig:exgal_examples} shows the source TGSSADR~J222603.3$+$172208, which was detected in our alternative search using the TGSS-ADR1 catalogue for prior positions. In this overlay, we also include archival C-configuration VLA data at 1.4\,GHz (project AS13, observed 1983 May 15) which was retrieved from the NRAO VLA Archive Survey (NVAS\footnote{The NVAS can currently be browsed through \url{http://archive.nrao.edu/nvas/}}) in an effort to identify a host object. However, as can be seen in the lower panel of Figure~\ref{fig:exgal_examples}, the polarised emission from this source measured at 200\,MHz appears offset from the emission at 1.4\,GHz. Given that it lies within the 150\,MHz TGSS-ADR1 Stokes $I$ emission, this polarised emission is likely associated with a steep-spectrum component of this complex radio source. We were unable to confidently identify a host object for TGSSADR~J222603.3$+$172208.

In \citetalias{Riseley2018}, we also identified a handful of radio galaxies where two polarised sources were detected, each associated with one of a pair of radio lobes. These were the radio galaxies PMN~J0351$-$2744 \citep[e.g.][]{Lenc2017}, ESO~422$-$G028 \citep[also known as MSH~05-22, e.g.][]{Subrahmanyan2008}, PKS~J0636$-$2036 \citep[e.g.][]{OSullivan2012} and PKS~0707$-$35 \citep[e.g.][]{Burgess2006}. We identify these four sources plus a further six physical pairs of polarised sources (`polarised doubles') in POGS ExGal, two examples of which are shown in Figure~\ref{fig:exgal_doubles}. Samples of such physical pairs of polarised sources can be used to provide constraints on the magnetised Cosmic Web \citep[e.g.][]{Vernstrom2019,OSullivan2020} but in order to derive meaningful constraints, we would require significantly higher surface density than is achieved with our MWA Phase I polarisation work.

These sources are (i) POGSII-EG-250 and POGSII-EG-251 (GLEAM~J080225$-$095823 and GLEAM~J080253$-$095822), which together form the radio lobes of PKS~B0800$-$09 \citep[e.g.][]{Bolton1968}, catalogued by \citet{White2020b} as `G4Jy 680', and (ii) POGSII-EG-265 and POGSII-EG-266 (GLEAM~J094633$-$132703 and GLEAM~J094739$-$134806), which comprise the hotspots associated with the giant radio galaxy `J0947$-$1338' \citep[e.g.][]{Kumicz2018}. The host galaxy of J0947$-$1338, AllWISE~J094708.00$-$133827.6, is associated with the compact radio source that falls on the $3\sigma$ GLEAM contour from the South-Eastern radio lobe (see the lower panel of Figure~\ref{fig:exgal_doubles}).

We note some disagreement in the literature regarding the host galaxy of PKS~B0800$-$09. The host currently adopted by the the NASA/IPAC Extragalactic Database (NED), marked by the green `x' in Figure~\ref{fig:exgal_doubles}, is AllWISE~J080239.90$-$095809.8 at $z=0.0892$ \citep[SDSS DR12;][]{Alam2015}. However, archival B/C configuration VLA data at 4.89\,GHz (project AJ~141, observed 1986 Oct 01) sourced from the NVAS reveal a probable radio core and one-sided jet associated with AllWISE~J080236.28$-$095739.9 \citep[$z=0.0699$;][]{Jones2009}, and so we suggest that this is in fact the correct host for PKS~B0800$-$09 (marked by a white `+'). We note that, as discussed by \cite{White2020b}, this was the host identification favoured by \citet{Schilizzi1975}. 

\begin{figure*}
\centering
\includegraphics[width=\textwidth]{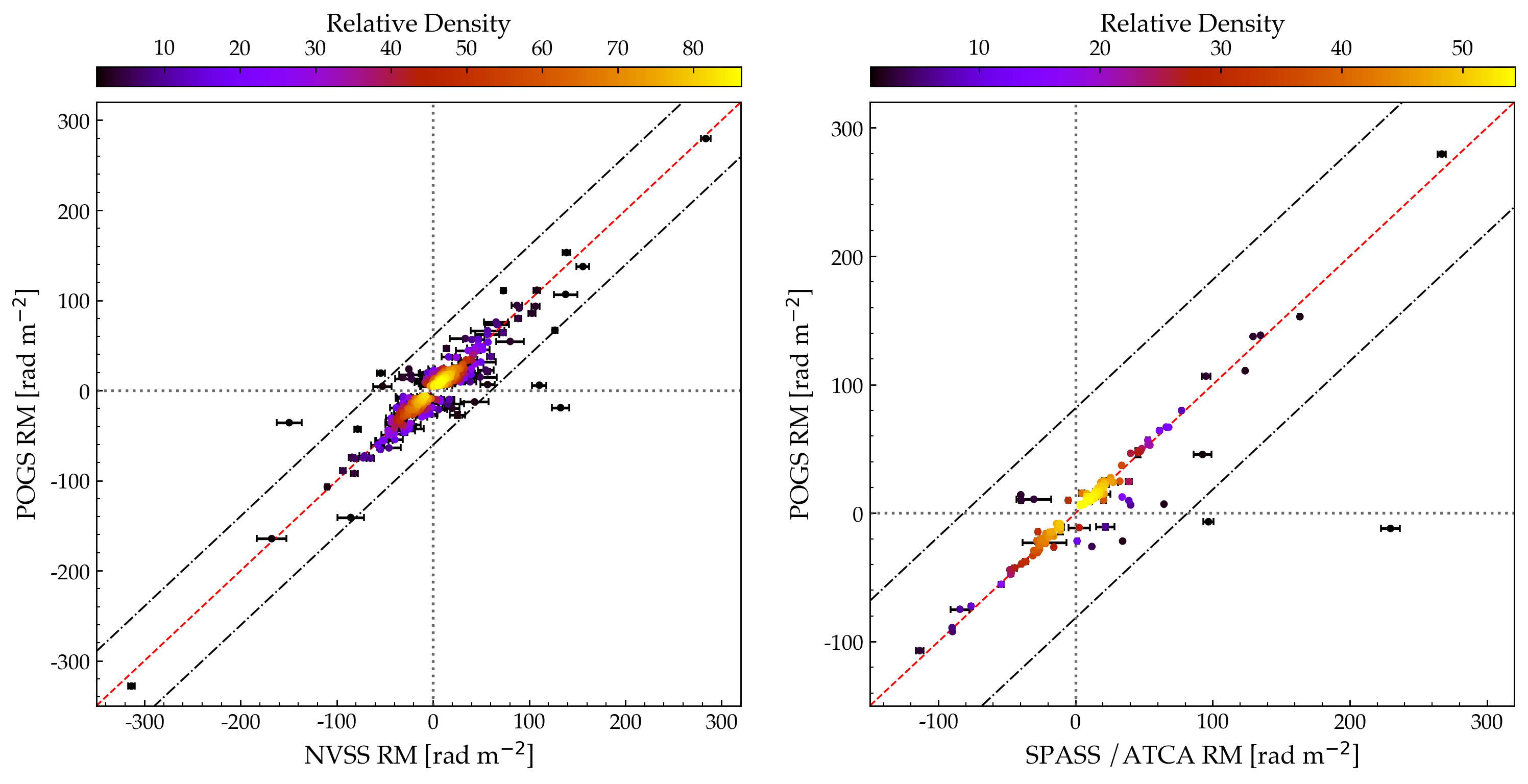}
\caption{Comparison of POGS ExGal RM with NVSS RM (\emph{left}) and S-PASS/ATCA RM (\emph{right}). Dashed red line denotes unity, dotted lines denote zero RM. The dot-dashed lines denote the $3\sigma$ scatter in the RM/RM plane. Individual markers are colourised according to the density in the RM/RM plane to assist the reader. Note that we show different axis ranges in each subplot.}
\label{fig:exgal_rm}
\end{figure*}

\subsection{RM Consistency}
There are two other RM catalogues in the literature which, between them, cover the full POGS survey area. The first of these is the 1.4\,GHz NVSS RM catalogue of \citet{Taylor2009}, which contains RMs for 37,543 sources at Declination $> -40\degree$, derived from the NVSS catalogue (which contains flux density measurements in both total intensity and linear polarisation; \citealt{Condon1998}). 

In the Southern sky, this is complemented by the 2.2\,GHz S-PASS/ATCA catalogue \citep{Schnitzeler2019}, containing RMs for 3,811 sightlines at Declination $<0\degree$ at a resolution of $2\times1$~arcmin. This catalogue is also uniquely suited to compare with POGS ExGal, as \citeauthor{Schnitzeler2019} used their broad-band data to investigate the potential for multiple RM components along their sightlines.

We cross-matched POGS ExGal with the NVSS-RM and S-PASS/ATCA-RM catalogues using a radius of 3~arcmin (our typical PSF) and visually inspected the results to confirm associations between polarised sources. We find a total of 286 POGS ExGal sources with NVSS-RM counterparts, and 124 with S-PASS/ATCA-RM counterparts. Figure~\ref{fig:exgal_rm} shows the comparison between our POGS ExGal RMs and the RMs from these reference catalogues.

From Figure~\ref{fig:exgal_rm}, the overwhelming majority of sources lie within $3\sigma$ of unity, suggesting that there is good agreement between POGS ExGal and the RMs derived at higher frequencies. Given that low-frequency RM studies are only sensitive to Faraday-thin sources, this suggests that the majority of these sources are also dominated by a single Faraday-rotating component.

Instead, we search for clear outliers away from the unity line. From Figure~\ref{fig:exgal_rm}, there are a number of clear outliers. The clearest outlier in the right panel of Figure~\ref{fig:exgal_rm} is GLEAM~J182331$-$705604, which has ${\rm{RM}}_{\rm{POGS}} = -11.86\pm0.68$~rad~m$^{-2}$ and ${\rm{RM}}_{\rm{S-PASS/ATCA}} = +229.66\pm6.94$~rad~m$^{-2}$. This source is identified in the peaked-spectrum sample of \citet{Callingham2017} and exhibits steep spectral behaviour above 1\,GHz with a flattening in the GLEAM band, so we suggest that this stark RM difference may be the result of different spectral index properties of two polarised emission components within this source. Polarisation observations across a wide frequency range, which would allow \emph{QU}-fitting, would be required to study this further.

There are also a handful of clear outliers in the left panel of Figure~\ref{fig:exgal_rm}. These sources are GLEAM~J100123$-$263720, GLEAM~J120533$-$263407, and GLEAM~J181835$+$240055. From inspection, we find that (i) all are associated with active galactic nuclei (AGN) and (ii) all remain unresolved by both the MWA and the NVSS. Unresolved polarised radio sources can exhibit complex spectropolarimetric behaviour, often showing different spectral indices in continuum and polarisation \citep[e.g.][]{Schnitzeler2019}. Alternatively, these may represent examples of RM time variability \citep[e.g.][]{Anderson2019}. Further polarisation observations across a broad frequency range would be required to study this behaviour further.

\subsection{Polarisation Properties of POGS ExGal}
\begin{figure}
\centering
\includegraphics[width=0.48\textwidth]{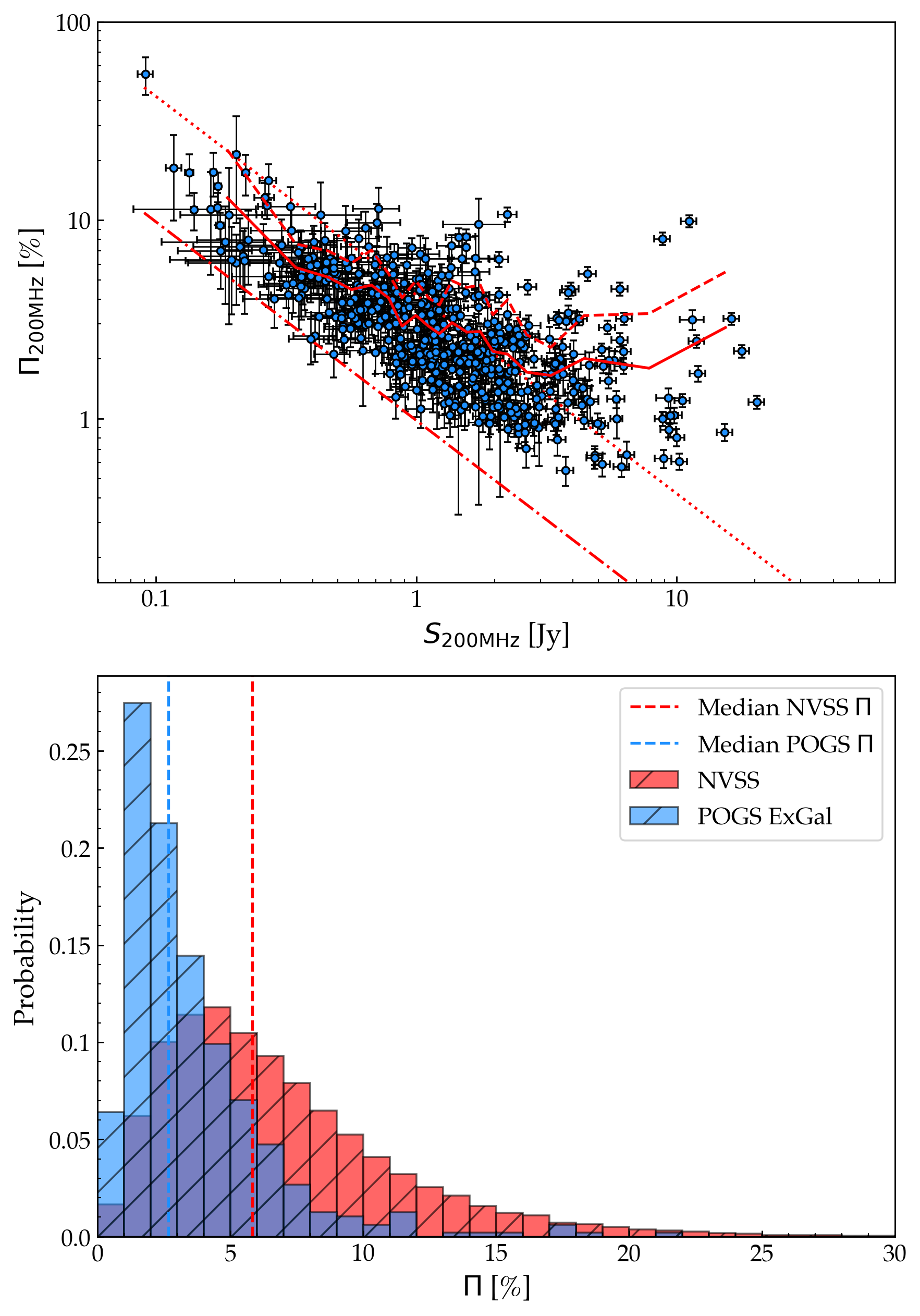}
\caption{Polarisation properties of extragalactic radio sources detected in POGS. {\emph{Upper:}} fractional polarisation as a function of 200\,MHz Stokes $I$ flux density. Solid (dashed) red lines denote the mean $({\rm{mean}}+1\sigma)$ $\Pi$, derived in adjacent bins of 25 sources. Dot-dashed and dotted lines denote the lower detectability bound for the Declination $-27\degree$ strip and the Declination $+18\degree$ strip, derived using a typical off-source rms noise of 1.4\,mJy and 6\,mJy, respectively. {\emph{Lower:}} fractional polarisation histogram for POGS ExGal (blue) and the NVSS RM catalogue (red). Dashed lines denote the median polarisation fraction for each sample. For POGS ExGal, this value is $\Tilde{\Pi}_{\rm{200\,MHz}} = 2.67\%$; for the NVSS RM catalogue, $\Tilde{\Pi}_{\rm{1.4\,GHz}} = 5.83\%$. }
\label{fig:exgal_pol}
\end{figure}

Figure~\ref{fig:exgal_pol} summarises the fractional polarisation $(\Pi)$ properties of POGS ExGal. The top panel shows the polarisation fraction at 200\,MHz $(\Pi_{\rm{200\,MHz}})$ as a function of Stokes $I$ flux density. Note that we have not merged the ten `polarised doubles' mentioned in the previous section, as these contribute a small fraction of our full sample. The central and lower panels show comparisons of our 200\,MHz fractional polarisation with that measured in the 1.4\,GHz NVSS RM catalogue. 

\subsubsection{Properties of the Full Sample}
In the top panel of Figure~\ref{fig:exgal_pol}, we also show the evolution in the mean fractional polarisation, $\Bar{\Pi}$, and the standard deviation, $\sigma({\Pi})$, as a function of Stokes $I$ flux density. These curves were derived by binning our sources according to Stokes $I$ flux density in adjacent bins of 25 sources, and then calculating $\Bar{\Pi}$ and $\sigma({\Pi})$ per bin.

From Figure~\ref{fig:exgal_pol}, two trends are visible in the $\Pi_{\rm{200\,MHz}}/S_{200\,{\rm{MHz}}}$ plane. The first trend is that there appears to be an inverse relationship between $\Pi_{\rm{200\,MHz}}$ and $S_{200\,{\rm{MHz}}}$. For the fainter source regime, where $S_{200\,{\rm{MHz}}} < 0.5$\,Jy, the mean fractional polarisation is $\Bar{\Pi}_{\rm{200\,MHz}}=7.8\%$, whereas for sources with $0.5\,{\rm{Jy}} < S_{200\,{\rm{MHz}}} < 3\,{\rm{Jy}}$, the mean is $\Bar{\Pi}_{200\,{\rm{MHz}}}=3.0\%$.

Similar trends have previously been observed at higher frequencies in large samples of polarised sources, where $\Pi$ rises from $\sim2.5\%$ at $S_{\rm{1.4\,GHz}}>10$\,mJy to $\sim15\%$ below 1\,mJy \citep[e.g.][]{Taylor2007,Subrahmanyan2010}. However, given that extragalactic radio sources typically exhibit low fractional polarisation, at fainter Stokes $I$ flux densities we are naturally biased towards sources with higher fractional polarisation. Indeed, by stacking polarised intensity images from the NVSS, \citet{Stil2014} showed that while $\Pi$ still increases with decreasing $S_{\rm{1.4\,GHz}}$, the slope is far more gradual than shown by previous surveys.

The second noticeable trend is that, for $S_{200\,{\rm{MHz}}} \gtrsim 3$\,Jy, the median polarisation fraction appears to flatten, before increasing above $\sim8$\,Jy. However, for this population of increasingly bright sources, any leakage of Stokes $I$ signal into the linear polarisation products (due to inaccuracies in the beam model or calibration errors, for example) can come to dominate any polarised signal. This means that again we are biased toward only sources with higher fractional polarisation, where the real polarised signal can be separated from instrumental leakage. Improvements in either the MWA beam model and/or advanced calibration techniques would be required to mitigate instrumental leakage further and probe the polarisation properties of this bright source population.

\subsubsection{Comparison with the NVSS RM Catalogue}
The lower panel of Figure~\ref{fig:exgal_pol} shows a histogram of fractional polarisation for POGS ExGal (484 sources) and the NVSS RM catalogue (37,543 sources), normalised according the sample size. For clarity, we show the region $\Pi<30\%$, which excludes only a single source from POGS ExGal and 10 sources from the NVSS RM catalogue.

The histogram suggests that the two populations exhibit different distributions in polarisation fraction. While both populations exhibit broadly similar scatter, with $\sigma(\Pi_{\rm{200\,MHz}})\sim3.62\%$ and $\sigma(\Pi_{\rm{1.4\,GHz}})\sim4.81\%$, the median polarisation fraction for POGS ExGal is $\Tilde{\Pi}_{\rm{200\,MHz}} = 2.67\%$, about half that for the NVSS RM catalogue, $\Tilde{\Pi}_{\rm{1.4\,GHz}} = 5.83\%$. We also note that the NVSS RM catalogue is likely biased at the low end of the fractional polarisation distribution, as \citet{Taylor2009} exclude sources with $\Pi_{\rm{1.4\,GHz}}<0.5\%$.

\subsection{POGS ExGal as a Probe of the Extragalactic Magnetised Universe}
For all our POGS ExGal sources, the observed RM combines contributions from multiple screens along the LOS. While we have already corrected for the Earth's ionosphere, there remains a contribution from the Milky Way's magnetised foreground: the Galactic RM. 

We have used POGS ExGal to probe the extragalactic RM component by deriving the residual rotation measure (RRM), defined as:
\begin{equation}
    {\rm{RRM}} \equiv {\rm{RM_{observed}}} - {\rm{RM_{Galactic}}}
\end{equation}
where ${\rm{RM_{Galactic}}}$ was measured using the all-sky Galactic RM reconstructed by \citet{Oppermann2015}. We note, however, that at Declination~$\leq-40\degree$, the Galactic RM reconstruction is based on a comparatively small sample of sources \citep[some $\sim900$;][]{Oppermann2015} compared to the sky at Declination~$>-40\degree$ (based on some $>40,000$ sources). 

\begin{figure*}
\centering
\includegraphics[width=0.9\textwidth]{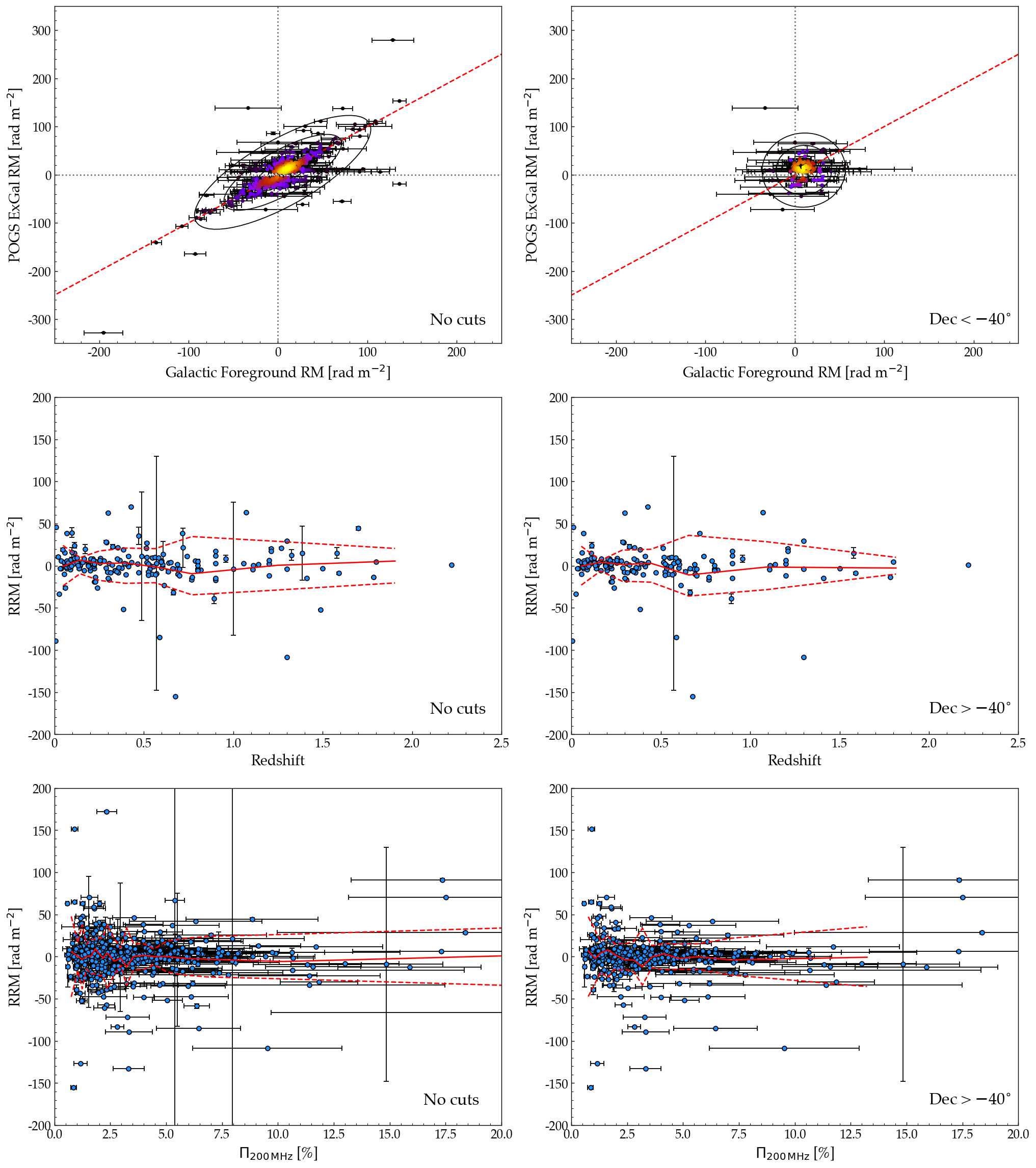}
\caption{\emph{Upper:} observed RM and Galactic RM for POGS ExGal sources. Different cuts are shown according to the inset. Dashed red line denotes unity, dotted lines denote zero RM. The dot-dashed ellipses denote the $1\sigma$, $2\sigma$ and $3\sigma$ scatter in the RM/RM plane. Individual markers are colourised according to the density in the RM/RM plane to assist the reader. \emph{Middle:} RRM as a function of redshift for the 179/484 POGS ExGal sources where a host with measured redshift could be found. \emph{Lower:} RRM as a function of polarisation fraction $(\Pi_{\rm{200\,MHz}})$ for POGS ExGal sources, whether or not a redshift could be found. For clarity, we show the region $\Pi_{\rm{200\,MHz}}\leq20\%$, excluding two sources. Different cuts on the population are indicated in the inset. Solid (dashed) red lines denote the mean $(\pm1\sigma)$ RRM in each plane, derived in adjacent bins of 25 sources.}
\label{fig:exgal_rrmrelations}
\end{figure*}

The upper row of Figure~\ref{fig:exgal_rrmrelations} shows the observed-RM/Galactic-RM plane for POGS ExGal sources, both across the whole sky (upper-left panel) and solely those at Declination~$\leq-40\degree$ (upper-right panel). From the upper-left panel of Figure~\ref{fig:exgal_rrmrelations}, it appears that there is broadly good agreement between POGS ExGal RM and Galactic RM, suggesting that the majority of the Faraday rotation we are measuring is being caused by the Galactic foreground. However, the relatively large number of outliers, far from the 1:1 line, suggests that there is a non-negligible extragalactic Faraday rotation component. It is also readily apparent that at Declination~$<-40\degree$, there is little correlation between POGS ExGal RM and Galactic RM. This lies below the NVSS Declination limit, so the significantly lower source density used in the Galactic RM reconstruction of \cite{Oppermann2015} results in larger uncertainty. Furthermore, the typical magnitude of source RMs in this region appears to be lower, thus increasing the fractional uncertainty. Future polarisation surveys with ASKAP (such as POSSUM, which includes a specific `RM-grid' goal; \citealt{Gaensler2010}) will be critical to fill in this sparsely-sampled (in polarisation) region of sky.

Finally, we note that a handful of POGS ExGal sources have extremely large uncertainties on their RRMs; this results from the large fractional uncertainty in Galactic RM along the LOS to these sources, which is readily apparent in the upper-right panel of Figure~\ref{fig:exgal_rrmrelations}. Thus, in the following sections, we will not only examine the whole sample together, but also a sub-sample at Declination~$>-40\degree$, where the higher density of sources used to reconstruct the foreground will mean that our RRMs are likely to be more reliable.

\subsubsection{RRM/redshift relation}
While it is well-established that there is no redshift evolution of RRM out to $z\sim4$ \citep[e.g.][]{Kronberg1977,Oren1995,Kronberg2008,Vernstrom2018}, the debate over the dependence (or lack thereof) of the variance in RRM with redshift has been more contentious. For example, the sample of $\sim300$ sources catalogued by \citet{Kronberg2008} suggested a significant increase in the variance of RRM with redshift, suggesting strong magnetic fields were present in galaxies in the relatively early Universe. However, from their larger sample of $\sim3650$ sources at $|b|\geq20\degree$, \citet{Hammond2012} find no significant evolution of RRM variance out to redshift $z\sim5.3$. 

We present the RRM/$z$ plane in the central row of Figure~\ref{fig:exgal_rrmrelations}, for two different cuts on our source population. Following \citet{Hammond2012}, we also show the mean and standard deviation of RM for this population as a function of redshift, derived using adjacent bins of 25 sources (although we note that we are using a smaller population in each bin as a result of our smaller overall population size).

From Figure~\ref{fig:exgal_rrmrelations}, there is no clear trend in either the mean or standard deviation of RRM with redshift, either for the full POGS ExGal population or the sources at Declination~$>-40\degree$. This is consistent with previous results derived from larger samples of polarised sources with known redshifts \citep{Bernet2012,Hammond2012,Vernstrom2018}. This suggests that the observed Faraday rotation is not internal to the sources, but is caused by an external screen.

\subsubsection{RRM/polarisation fraction relation}
In \citetalias{Riseley2018}, we performed an initial study of the relation between RM and polarisation fraction for our extragalactic source population, following the work of \citet{Hammond2012}. We have revisited this with our larger POGS ExGal sample, the result of which is shown in the lower panels of Figure~\ref{fig:exgal_rrmrelations}. Note that for clarity, we show the region $\Pi_{200\,\rm{MHz}} \leq 20\%$; this region excludes only two sources from POGS ExGal. We note that the sources which exhibit the highest polarisation fraction also show the largest measurement uncertainty. However, this is to be expected -- these are among the faintest sources in our catalogue, so can only be detected because they have a high polsarisation fraction. 

From Figure~\ref{fig:exgal_rrmrelations}, we see no significant evolution of RRM with $\Pi$, consistent with previous studies. Interpreting the standard deviation of RRM is slightly more complex. Toward higher fractional polarisation, we are relatively limited by small number statistics, so we focus on the region $\Pi_{\rm{200\,MHz}} \lesssim 8\%$, which contains some $\sim95\%$ of our sources (from Figure~\ref{fig:exgal_pol}, middle panel). In this region, we see roughly a 50\% decrease in the standard deviation of RRM (from $\sim48$~rad~m$^{-2}$ to $\sim24$~rad~m$^{-2}$).

Using the NVSS RM catalogue, \citet{Hammond2012} found a similar trend -- an anticorrelation between RRM variance and fractional polarisation (their Figure 19). However, as recently shown by \cite{Ma2019a,Ma2019b} toward lower fractional polarisation ($\Pi_{\rm{1.4\,GHz}}\lesssim1\%$) the NVSS RM catalogue may be contaminated by off-axis leakage effects, which introduces extra RM uncertainties of about 13.5~rad~m$^{-2}$ \citep[see][]{Ma2019b}. Nevertheless, when taking only sources with $\Pi_{\rm{1.4\,GHz}}\gtrsim2\%$, a similar $\sim60\%$ decrease in RRM variance is visible in Figure 19 of \citet{Hammond2012}, from $\sim25$~rad~m$^{-2}$ to $\sim10$~rad~m$^{-2}$.

As discussed by \cite{Hammond2012}, this anticorrelation likely has an astrophysical origin, resulting from some depolarisation mechanism \citep[e.g.][]{Burn1966,Sokoloff1998}. Given the long wavelength of our observations, $\lambda^2 \gg 1$, we are extremely sensitive to depolarisation, so it is likely that this is the cause of the behaviour we observe.

We do not consider it likely that we suffer from significant bandwidth depolarisation, as we are sensitive to RMs up to $|{\rm{RM}}|\sim1100$~rad~m$^{-2}$, whereas all our sources have $|{\rm{RM}}|\lesssim330$~rad~m$^{-2}$. Faraday depth depolarisation could play a role, as our observations are insensitive to structures thicker than around $1.9$~rad~m$^{-2}$. However, given that we observe no evolution of RRM with redshift (Figure~\ref{fig:exgal_rrmrelations}, central panels), it is likely that the Faraday rotation we are measuring is occurring along the LOS to these sources, whereas Faraday-thick structures arise from a medium that both emits and rotates polarised emission. 

It has been demonstrated that low-frequency observations may recover the outer `skins' of a Faraday-thick structure \citep[e.g.][]{VanEck2018}. In these circumstances, the two skins would appear as separate Faraday-thin peaks in the RM spectrum (whereas none of our sources exhibited signs of multiple RM peaks) and would suffer significant depolarisation. As such, we also consider the Faraday depth depolarisation explanation unlikely. Thus the remaining explanation is beam depolarisation, whereby RM variations on scales smaller than the PSF cause the polarisation angle to rotate, decreasing the observed polarisation. Given that our study uses exclusively long-wavelength data, and the Faraday rotation is a function of wavelength-squared, we will be extremely sensitive to such fluctuations.

While all of our sources remain unresolved with our moderate resolution of around $3-8$ arcminutes, the effect of beam depolarisation has been observed in unresolved sources \citep[e.g.][]{Haverkorn2008}. Considering that Figure~\ref{fig:exgal_rrmrelations} shows no redshift evolution of RRM (either magnitude or standard deviation), this RM variation must occur along the LOS between a source and the observer. A natural explanation for this could be small-scale variation (i.e. at or below the scale of our PSF) in the Galactic foreground RM, which a number of sensitive, small-area studies have shown to be significant \citep[e.g.][]{Mao2010,Wolleben2010,Stil2011,Sun2015,Anderson2015}.

As shown by \cite{Stil2011}, the standard deviation in RM on angular scales at or below $1\degree$ (the typical sampling of sources used by \citealt{Oppermann2015} to reconstruct the Galactic foreground RM) is $\sim12-17$~rad~m$^{-2}$ for Galactic latitudes $|b| \geq 20\degree$. Selecting only those POGS ExGal sources at Declination $\geq -40\degree$ and $|b| \geq 20\degree$ and subtracting (in quadrature) a typical $15$~rad~m$^{-2}$ from our observed RRM variance, we find a persistent excess RRM variance of $\sim10-25$~rad~m$^{-2}$. This is broadly consistent with the predicted RM due to the halo of a Milky-Way-like galaxy \citep[e.g.][]{Mao2010,Mao2012}.

\section{Pulsars}\label{sec:psr}
\subsection{Known Pulsars}
Our catalogue contains 33 known pulsars from the ATNF Pulsar Catalogue\footnote{\url{https://www.atnf.csiro.au/research/pulsar/psrcat/}} \citep[hereafter ATNF psrcat;][]{Manchester2005}. For six of these pulsars, we provide the first recorded RMs. We measure linearly-polarised flux densities between 17\,mJy and 1.7\,Jy, with RMs between $-185.99$~rad~m$^{-2}$ and $+150.74$~rad~m$^{-2}$. The mean RM uncertainty is $0.22$~rad~m$^{-2}$, and the worst-case RM uncertainty is $2.84$~rad~m$^{-2}$.

An excerpt from our catalogue is presented in Table~\ref{tab:known_psr}, along with ancillary information sourced from the GLEAM Pulsar Catalogue \citep{Murphy2017} and the ATNF psrcat. We also present the RM spectra for all 33 known pulsars in Figure~C1 (Appendix~\ref{sec:appendix_c}). The properties of some pulsars in our catalogue are not listed in either of these catalogues, and were sourced from alternative references. These are listed here for completeness:
\begin{itemize}
    \item \textbf{PSR~J1747$-$4036}: this pulsar was not detected by \citet{Murphy2017}, but a compact radio source (GLEAM~J174749$-$403650) was catalogued within 30~arcsec by \citet{HurleyWalker2019}. We suggest that they are the same source, so we quote that object's 200\,MHz flux density.
    \item \textbf{PSR~J0509$+$0856}: the literature RM and DM for this pulsar are quoted from \citet{Martinez2019}.
\end{itemize}

There are two pulsars in our catalogue that are part of a pulsar binary system (J0737$-$3039A) or are located in a globular cluster that is known to host multiple pulsars (J1824$-$2452A, located in globular cluster M28). For the pulsar binary, the companion pulsar (J0737$-$3039B) precessed out of our line of sight in 2008 March and is expected to reappear around 2035, due to relativistic spin (or geodetic) precession \citep[e.g.][]{Perera2010}. 

Globular cluster M28 is known to host multiple pulsars, so it is possible that we are detecting an amalgamation of polarised emission from a number of these with our moderate resolution. However, we consider this unlikely for a number of reasons. Firstly, J1824$-$2452A is the dominant pulsar in M28, with a 1.4\,GHz flux density of $\sim2.3$\,mJy \citep{Dai2015}, whereas the second-brightest pulsar, J1824$-$2452C, has a 1.4\,GHz flux density of $\sim0.17$\,mJy. While the low-frequency spectral behaviour of the other pulsars is not known, with a spectral index $\alpha=-3.2\pm0.1$ \citep{Murphy2017}, `pulsar A' likely dominates the low-frequency continuum. Secondly, we do not detect any additional peaks in the RM spectrum of this source that might indicate other pulsars within our PSF. Thirdly, if we detected emission from other pulsars within our PSF at similar RM to pulsar A, it could result in a broadening of the apparent RM peak. However, for J1824$-$2452A, the width of the RM spectrum peak is consistent with that of other isolated pulsars in our catalogue. 

Thus, for both J0737$-$3039A and J1824$-$2452A, we assume that the emission we are detecting is associated with `pulsar A' of each system.

\begin{sidewaystable*}

\centering
\small
\caption{Sample rows from POGS PsrCat, showing only columns that vary by source. Note that, for display purposes, unfilled entries are marked with a `$-$'. \label{tab:known_psr}}
\begin{tabular}{|crrrrrrrrrr|l}
\hline
POGS ID & RA & Dec & $l$ & $b$ & Pos.Err. & RM & $S_{\rm{200\,MHz}}$ & $\alpha$ & $P_{\rm{200\,MHz}}$ & $\Pi_{\rm{200\,MHz}}$ \\
& $[\degree]$ & $[\degree]$ & $[\degree]$ & $[\degree]$ & $[\degree]$ & [rad m$^{-2}$] & [Jy] & & [Jy] & [\%] \\
\hline
POGSII-PS-001 & $8.54947$ & $-7.36550$ & $110.45536$ & $-69.81854$ & $0.00073$ & $10.91\pm0.08$         & $0.29\pm0.01$ & $-1.40\pm0.10$ & $0.057\pm0.004$         & $20.0\pm2.0$ & ... \\
POGSII-PS-002 & $73.15290$ & $-17.98862$ & $217.08130$ & $-34.07673$ & $0.00046$ & $13.11\pm0.07$         & $0.10\pm0.01$ & $-0.30\pm0.10$ & $0.035\pm0.002$         & $36.0\pm4.0$ & ... \\
POGSII-PS-003 & $77.35735$ & $8.93894$ & $192.49585$ & $-17.91912$ & $0.00081$ & $44.29\pm0.14$         & $-$ & $-$ & $0.033\pm0.006$         & $-$ & ... \\
POGSII-PS-004 & $97.70711$ & $-28.57466$ & $236.94861$ & $-16.75551$ & $0.00010$ & $46.65\pm0.02$         & $0.46\pm0.00$ & $-1.20\pm0.10$ & $0.211\pm0.003$         & $46.0\pm1.0$ & ... \\
POGSII-PS-005 & $114.46681$ & $-30.65502$ & $245.23158$ & $-4.49932$ & $0.00038$ & $121.19\pm0.08$         & $0.06\pm0.01$ & $-2.60\pm0.10$ & $0.029\pm0.002$         & $45.0\pm8.0$ & ... \\
... & ... & ... & ... & ... & ... & ... & ... & ... & ... & ... & ... \\
POGSII-PS-029 & $294.91691$ & $21.57361$ & $57.50350$ & $-0.29932$ & $0.00029$ & $8.54\pm0.03$         & $-$ & $-$ & $0.571\pm0.012$         & $-$ & ... \\
POGSII-PS-030 & $312.14848$ & $-16.28139$ & $30.51171$ & $-33.07743$ & $0.00056$ & $-9.59\pm0.13$         & $0.17\pm0.01$ & $-0.60\pm0.20$ & $0.027\pm0.003$         & $16.0\pm2.0$ & ... \\
POGSII-PS-031 & $328.80472$ & $-31.31002$ & $15.85348$ & $-51.57525$ & $0.00067$ & $13.67\pm0.21$         & $0.05\pm0.01$ & $-2.00\pm0.10$ & $0.022\pm0.004$         & $48.0\pm11.0$ & ... \\
POGSII-PS-032 & $340.40949$ & $-52.60757$ & $337.47020$ & $-54.92023$ & $0.00104$ & $12.64\pm0.11$         & $0.06\pm0.01$ & $-1.30\pm0.10$ & $0.025\pm0.004$         & $41.0\pm10.0$ & ... \\

\hline
\end{tabular}
\bigskip

\begin{tabular}{r|rrrllc|rrr|}
\hline
 & & & & & & & \multicolumn{3}{c|}{Reference Values} \\
 & Bmaj & Bmin & BPA & DataID & Notes & $n\alpha$ & period & DM & RM \\
 & $[\degree]$ & $[\degree]$ & $[\degree]$ & & & & [s] & [pc cm$^{-3}$] & [rad m$^{-2}$] \\
\hline
... & 0.06500 & 0.05400 & 137.4 & J0034$-$0721 & - & 2 &         0.9430 & 10.922 & $9.89\pm0.07$ \\
... & 0.05800 & 0.04900 & 181.2 & J0452$-$1759 & - & 2 &         0.5489 & 39.903 & $13.80\pm0.70$ \\
... & 0.06500 & 0.05500 & 180.6 & J0509$+$0856 & MSP,binary & 0 &         0.0041 & 38.318 & $42.40\pm0.60$ \\
... & 0.05300 & 0.04700 & 268.3 & J0630$-$2834 & - & 2 &         1.2444 & 34.425 & $46.53\pm0.12$ \\
... & 0.05300 & 0.04700 & 268.3 & J0737$-$3039A & MSP,double pulsar & 1 &         0.0227 & 48.920 & $112.30\pm1.50$ \\
... & ... & ... & ... & ... & ... & ... & ... & ... & ...  \\
... & 0.09900 & 0.07800 & 177.0 & J1939$+$2134 & MSP & 0 &         0.0016 & 71.024 & $6.70\pm0.60$ \\
... & 0.06300 & 0.05400 & 270.5 & J2048$-$1616 & - & 2 &         1.9616 & 11.456 & $-10.00\pm0.07$ \\
... & 0.05100 & 0.04700 & 352.5 & J2155$-$3118 & - & 1 &         1.0300 & 14.850 & $21.00\pm3.00$ \\
... & 0.05500 & 0.05200 & 281.2 & J2241$-$5236 & MSP,binary & 1 &         0.0022 & 11.411 & $14.00\pm6.00$ \\
... & 0.06400 & 0.05300 & 181.3 & J2256$-$1024 & MSP,binary & 0 &         0.0023 & 13.800 & $-$ \\
\hline
\end{tabular}

\end{sidewaystable*}

\begin{figure}
\centering
\includegraphics[width=0.48\textwidth]{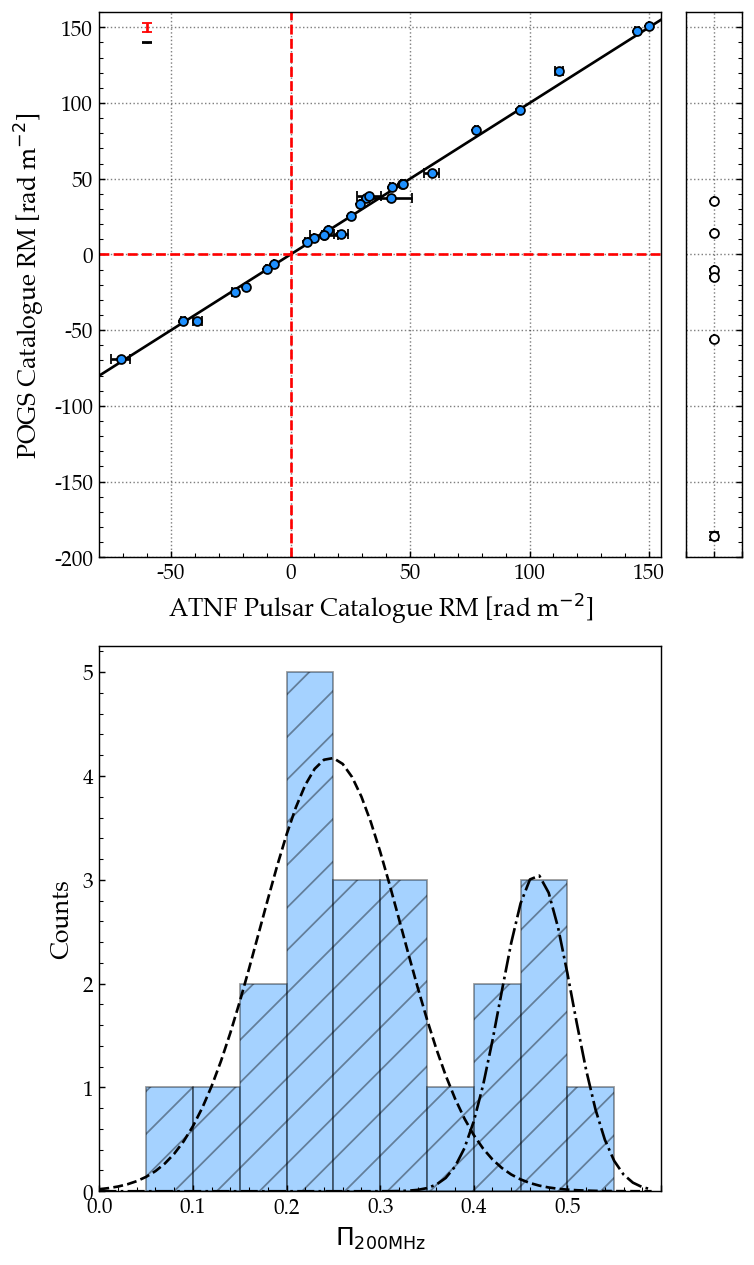}
\caption{Polarisation properties of known pulsars detected in POGS. {\emph{Top panel:}} Comparison of rotation measures (RMs) for known pulsars in the ATNF psrcat and POGS PsrCat at 200\,MHz. Pulsars without RMs in the ATNF psrcat are shown as empty markers in the right-hand panel. Red dashed lines mark zero RM; black line denotes unity. The median and worst-case measurement uncertainties from these pulsars in each catalogue are indicated respectively by the black and red error symbols in the upper-left quadrant. {\emph{Bottom panel:}} Histogram of 200\,MHz fractional polarisation for the 22/33 pulsars with continuum image-plane detections in GLEAM survey data \protect\citep{Murphy2017,HurleyWalker2019}. Dashed and dot-dashed lines represent Gaussians fitted to the population, with typical polarisation fractions of $\Pi_{200\,\rm{MHz}}=24.7\pm1.4\%$ and $46.7\pm1.4\%$, respectively.}
\label{fig:known_psr}
\end{figure}

In Figure~\ref{fig:known_psr} we present properties of the pulsars we have detected in linear polarisation at 200\,MHz. The top panel presents a comparison of our 200\,MHz RMs with those measured in the literature; our RMs are typically in very good agreement with the values determined at other frequencies. We also note that our median uncertainty $(0.12$~rad~m$^{-2})$ represents a $40\%$ improvement on the median uncertainty for sources with known RMs $(0.20$~rad~m$^{-2})$. Based on our relatively small sample, this suggests that there is no frequency-dependence of the RMs, and therefore no contribution to the observed RM from the pulsar magnetosphere, at least to the precision of our measurements, in agreement with previous studies \citep[e.g.][and references therein]{Sobey2019}.

The lower panel of Figure~\ref{fig:known_psr} shows a histogram of the polarisation fraction for POGS PsrCat pulsars that have continuum image-plane detections in GLEAM survey data \citep{Murphy2017,HurleyWalker2019}. Note that only three of these pulsars are known to exhibit long-term variability, likely due to magnetospheric emission mode changes (PSRs J0034$-$0721 and J0828$-$3417; e.g. \citealt{McSweeney2017,Esamdin2005}) or refractive interstellar scintillation (PSR~J0630$-$2834; e.g. \citealt{Bell2016}). Our use of GLEAM data means that both continuum and polarisation data will have been taken in the same epoch.

The histogram appears to exhibit a bimodal distribution. We performed a one-sample Kolmogorov-Smirnov (K-S) test on our measured $\Pi_{200\,\rm{MHz}}$ values to determine the likelihood that this apparent bimodality was due to random chance. Our test yielded a K-S statistic of 0.53 and associated $p$-value of $1.67\times10^{-6}$, so we can confidently reject the null hypothesis that our polarisation fraction measurements are drawn from the same normal distribution.

We thus fitted a pair of Gaussians to describe the population. We find our catalogue contains a dominant population of pulsars with typical polarisation fraction and variance of $\Pi_{200\,\rm{MHz}}=0.25$ and $\sigma_{\Pi}=0.08$, and a smaller population with $\Pi_{200\,\rm{MHz}}=0.47$ and $\sigma_{\Pi}=0.04$. None of the pulsars in this latter group show any common traits in terms of DM, spin period, location or spectral index properties. Note that, as a general rule, pulsar emission tends to increase in polarisation fractions towards low frequencies, thought to be due to the pulsar magnetospheric radio emission mechanism \citep[e.g.][]{Johnston2008,Noutsos2015}, with some exceptions \citep[e.g.][]{Xue2019}.

\subsubsection{Comparison with VCS data}
\citet{Xue2017} present a catalogue of 50 pulsars detected using the MWA's Voltage Capture System \citep[VCS;][]{Tremblay2015} at 185\,MHz. Of the 33 pulsars that we detect, 13 are common to the catalogue of \citeauthor{Xue2017}. Recently, \citet{Xue2019} have demonstrated that the VCS can be reliably calibrated in full-polarisation by observing two known, bright and strongly-polarised pulsars: J0742$-$2822 and J1752$-$2806. Both of these are in our catalogue, allowing additional cross-verification between our imaging data and MWA-VCS observations.

For PSR~J0742$-$2822, we cannot compare polarisation fraction measurements with those found by \citet{Xue2019}, as this source was not detected by \citet{Murphy2017}. However, we can compare RMs. From our image-plane data, we find an ${\rm{RM}}=+150.74\pm0.02$~rad~m$^{-2}$ at 200\,MHz; the VCS data indicate an ${\rm{RM}}=+150.975\pm0.097$~rad~m$^{-2}$ at 179\,MHz. For PSR~J1752$-$2806, our image-plane RM is ${\rm{RM}}=+95.74\pm0.08$~rad~m$^{-2}$; \citet{Xue2019} determine an RM of ${\rm{RM}}=+95.871\pm0.078$~rad~m$^{-2}$ from their VCS data at 154\,MHz. While these discrepancies are relatively significant (respectively $2.4\sigma$ and $1.7\sigma$), this is likely due to differences in the ionospheric RM correction applied to our different datasets. With the naturally high precision RMs that can be determined by our long-wavelength data, uncertainties in the ionospheric RM correction become the dominant contribution to the overall measurement uncertainty, so we consider our results broadly consistent with the VCS data.

We also note that PSR~J0742$-$2822 is known to change emission mode on the timescale of $\sim95$ days \citep{Keith2013}. The observations of \citet{Xue2019} were performed in 2016, whereas the GLEAM observations from which our catalogue was compiled were performed in 2013, so it is plausible that the data were taken when this pulsar was in different emission modes. This is not expected to cause a discrepancy in the RM, since we expect this effect to be the result of the ISM propagation, with no/negligible contribution from the relativistic electron-positron plasma in the pulsar magnetosphere \citep[e.g.][]{Melrose2017}, particularly at low observing frequencies \citep[e.g.][]{Wang2011,Noutsos2015}.

For PSR~J1752$-$2806, we measure a polarisation fraction of $\Pi_{200\,\rm{MHz}} = 7.0\pm1.3\%$. This is about half that determined for this pulsar from VCS data, and significantly discrepant with the general trend exhibited in measurements from the literature \citep[Figure 10 of][]{Xue2019}. The cause of this is likely to be the polarisation angle discontinuity across the pulse profile of PSR~J1752$-$2806 (Figure 11 of \citealt{Xue2019}), which will cause depolarisation when averaging over the pulse profile in the image domain.

\subsubsection{Consistency of Our Image-Plane Pulsar Measurements}
PSR~J1752$-$2806 was only detected in a single GLEAM epoch, so we cannot determine the cause of this discrepancy between our measured polarisation fraction and that determined by \citet{Xue2019}. However, PSR~J0742$-$2822 was in detected in two separate epochs of GLEAM observations, during observations on 2013 Nov. 25 and 2014 Mar. 03. Both epochs were observations of the Declination~$-27\degree$ strip, meaning that the primary beam response to this pulsar should be consistent.

The RMs were broadly consistent between epochs: in the first epoch, ${\rm{RM}}=+150.67\pm0.01$~rad~m$^{-2}$; for the second, ${\rm{RM}}=+150.80\pm0.02$~rad~m$^{-2}$. However, the polarised flux density measurements are inconsistent for these two epochs: $284.7\pm3.8$\,mJy and $241.5\pm4.1$\,mJy. These observing epochs are separated by 105 days, so these differences could be caused by a change in emission mode in the intervening period. However, as mentioned previously, our observations may suffer from some second-order ionospheric effect that may explain this discrepancy---this was observed in long-track observations of a known pulsar during LOFAR polarisation commissioning (priv. comm. LOFAR MKSP). Ten pulsars in our catalogue were detected in multiple epochs; all exhibit some level of apparent variability in polarised flux density between epochs (between $\sim 5\%$ and $\sim250\%$). However, three of these pulsars were each detected in four epochs, and our measured polarised flux densities are consistent in a sub-set of epochs. As with the POGS ExGal sources, we also note that our RMs were consistent between all multi-epoch detections.

Interstellar scintillation may also explain this phenomenon; when comparing single-epoch VCS data with the GLEAM Pulsar Catalogue, \citet{Xue2017} noted variations in continuum flux density measurements for many pulsars common to both catalogues. These differences ranged from the $\sim1.5\%$ level to the $\sim85\%$ level. The LOS to PSR~J0742$-$2822 passes through the Gum Nebula, which has been shown to be responsible for significant turbulence along this LOS \citep{Johnston1998}.

\subsubsection{Relation to Galactic Structure}
This new era of low-frequency polarimetry has allowed observers to determine very precise RMs for many known pulsars, unlocking a new window into probing the Galactic magnetic field. For example, \citet{Sobey2019} used beam-formed LOFAR High-Band Antenna (HBA) detections of 137 known pulsars to study the 3D Galactic halo magnetic field. For pulsars at distance $d$, with known RM and dispersion measure (DM), the ratio between RM and DM can be used to estimate the electron-density-weighted average magnetic field strength along the LOS, via:
\begin{equation}\label{eq:rmdm}
    \langle B_{\|} \rangle = \frac{ \int^0_d n_{\rm{e}} B_{\|} {\rm{d}}l }{ \int^d_0 n_{\rm{e}} {\rm{d}}l } = 1.232\,\upmu{\rm{G}}\left(\frac{{\rm{RM}}}{{\rm{rad~m}}^{-2}}\right)\left(\frac{{\rm{DM}}}{{\rm{pc~cm}}^{-3}}\right)^{-1}
\end{equation}
where $n_{\rm{e}}$ is the electron density (cm$^{-3}$) and ${\rm{d}}l$ is the unit path length along the LOS. We note that this relation relies on the assumption that there is no correlation between electron density and magnetic field strength \citep[e.g.][]{Beck2003}. If there is positive correlation between these quantities, RM will be enhanced and thus Equation~\ref{eq:rmdm} will over-estimate $\langle B_{\|}\rangle$; likewise, anti-correlation would result in an under-estimate of $\langle B_{\|}\rangle$.

\begin{figure}
\includegraphics[width=0.45\textwidth]{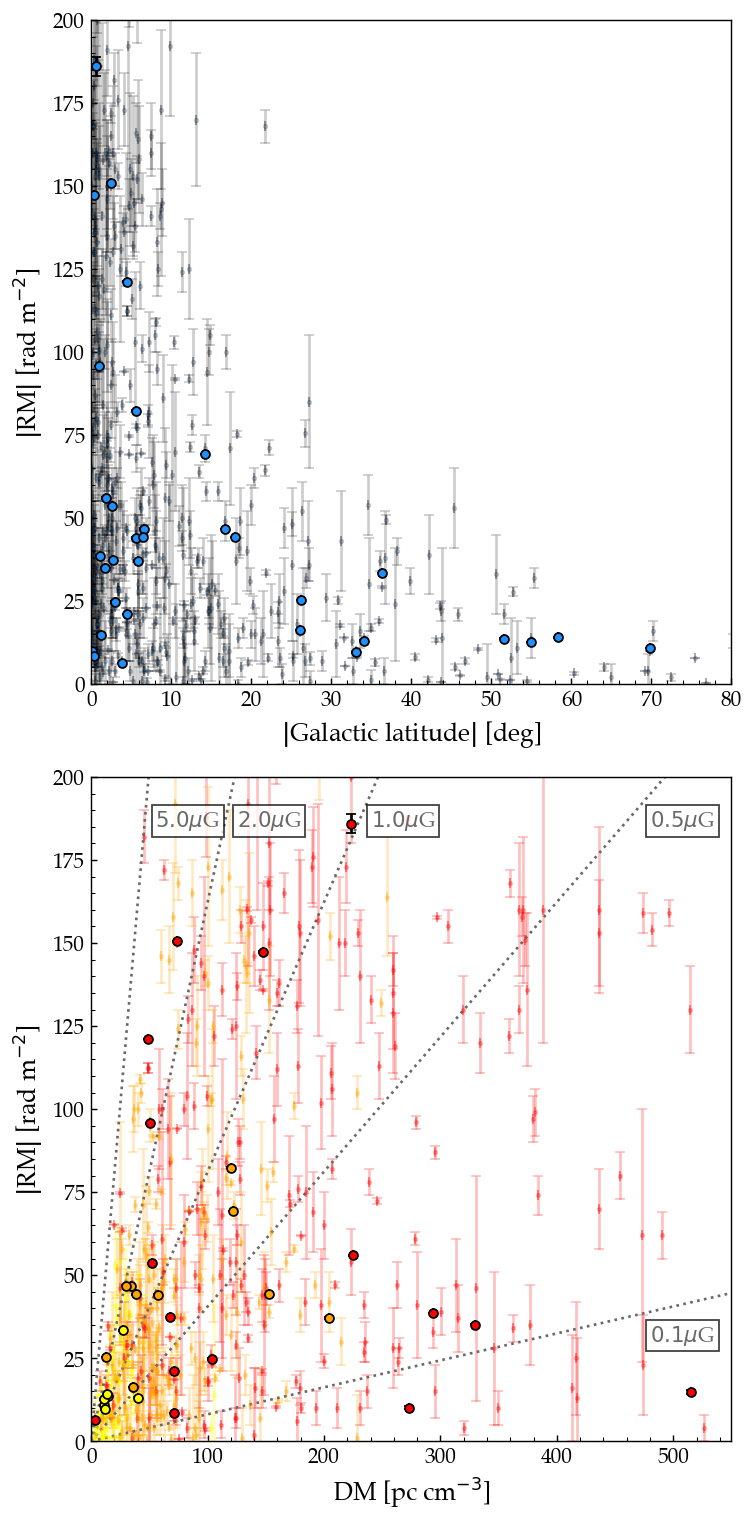}
\caption{Polarisation properties of the 33 pulsars in POGS PsrCat (filled symbols) and the 686 pulsars in the ATNF psrcat that have both RM and DM values (small, semitransparent symbols). \emph{Top panel:} absolute RM (i.e. $|{\rm{RM}}|$) as a function of absolute Galactic latitude (i.e. $|b|$). \emph{Bottom panel:} relation between absolute RM and DM. Colours represent different Galactic latitudes above and below the plane: red denotes $|b|\leq5\degree$, orange indicates $5<|b|<30\degree$ and yellow denotes $|b|\geq30\degree$. Gray lines show constant $|\langle B_{\|} \rangle|$ derived according to Equation~\ref{eq:rmdm}.}\label{fig:known_psr_gal}
\end{figure}

While our sample is much smaller, and based on imaging (rather than tied-array) data, we can also attempt to investigate this using catalogue DMs from the ATNF psrcat. Since the uncertainty in RM is usually the dominant source of uncertainty in $|\langle B_{\|} \rangle|$, our low-frequency RM measurements provide more accurate $|\langle B_{\|} \rangle|$.

The top panel of Figure~\ref{fig:known_psr_gal} shows the relation between absolute RM value and absolute Galactic latitude, for POGS PsrCat entries (filled points) and ATNF psrcat entries (open markers). Two clear trends are visible. Firstly, absolute RM decreases with increasing absolute Galactic latitude. Secondly, the scatter in absolute RM is significantly greater at low Galactic latitude. For POGS pulsars at $|b|\leq5\degree$, $\sigma_{|{\rm{RM}}|} = 56.5$~rad~m$^{-2}$, whereas for $|b|>5\degree$, $\sigma_{|{\rm{RM}}|} = 20.7$~rad~m$^{-2}$. 

The bottom panel of Figure~\ref{fig:known_psr_gal} also shows the variation of absolute RM with DM for the same population, for $|{\rm{RM}}| \leq200$~rad~m$^{-2}$ and $|{\rm{DM}}| \leq550$~pc~cm$^{-3}$. Markers are colourised according to Galactic latitude, with low-, medium-, and high-latitude pulsars indicated respectively in red, orange, and yellow.

Broadly-speaking, the lower panel of Figure~\ref{fig:known_psr_gal} indicates that pulsars located toward higher Galactic latitudes tend to have lower values of RM and DM, and are distributed across a fairly narrow range of Galactic magnetic field strengths, $|\langle B_{\|}\rangle| \leq 1.5\,\upmu$G. Pulsars toward lower Galactic latitudes tend to have larger RM and/or DM values, yet are also distributed across a fairly small range of Galactic magnetic field, $|\langle B_{\|}\rangle| \leq 3.1\,\upmu$G. This is largely consistent with the LOFAR results presented by \citet{Sobey2019}.

For two pulsars, J1601$-$5244 and J1851$-$0241, their position in the DM$/|{\rm{RM}}|$ plane suggests very low magnetic field strength: from Equation~\ref{eq:rmdm}, $|\langle B_{\|}\rangle| = 0.035\,\upmu$G for J1851$-$0241 and $0.046\,\upmu$G for J1601$-$5244. However, their locations towards the Galactic centre, proximity to the Galactic plane, and DM distances ($\sim$4.09 and 7.92 kpc; \citealt{Yao2017}) could also mean that their emission traverses through the magnetic field reversal(s) within the Galactic disk \citep[e.g.][]{VanEck2011}.

From Figure~\ref{fig:known_psr_gal} it appears that, similarly to \citet{Sobey2019}, we detect very few pulsars with both large DM and large $|{\rm{RM}}|$. This is converse to the population of pulsars in the ATNF PsrCat, which are largely uniformly distributed in DM$/|{\rm{RM}}|$-space.

We also note that our image-domain search retains sensitivity to pulsars with large DMs. The highest-DM pulsar for which we find a RM has a DM of 515~pc~cm$^{-3}$, which is significantly larger than the highest-DM pulsars typically detected in low-frequency, time-domain, beamformed data. For example, from the incoherent beamformed data presented by \citet{Xue2017}, the highest-DM pulsar has a DM of 147.45~pc~cm$^{-3}$. The highest-DM for which \citet{Sobey2019} detected an RM was 161~pc~cm$^{-3}$; the highest-DM pulsar yet detected by LOFAR is around 217~pc~cm$^{-3}$ \citep{Pilia2016}. A number of effects may account for this, such as dispersion smearing (for incoherent dedispersion) or scattering effects (that may smear the pulse profile over a larger number of profile bins); spectral index effects (particularly towards areas of increased sky temperature from the Galactic foreground where a pulsar's spectral index may be unfavourably small); propagation effects (e.g. scintillation) or instrumental effects (e.g. beam jitter). For further discussion of such effects, see for example \cite{Kondratiev2016} or \cite{Bilous2016}.

\subsection{New Pulsar Candidates}
In \citetalias{Riseley2018} we attempted to search for new pulsar candidates among our nominally extragalactic source population. We applied three selection criteria: a source must (a) be compact at the resolution of GLEAM and the NVSS, (b) exhibit a high polarisation fraction $(\gtrsim10\%)$ and (c) have a steep radio spectrum $(\alpha \lesssim -1)$. By applying these criteria, we found a single candidate pulsar among the 80 sources detected in polarisation (excluding the known pulsar PSR~B0628$-$28).

However, statistical samples have shown that pulsars may exhibit a wide range of spectral properties. For example, \citet{Bilous2016} showed that while 75\% of their sample were well-described by a single power-law spectrum, this strongly depends on the availability of multi-frequency flux density measurements. Furthermore, the index of this power-law varied significantly, with single-component spectra having $-3\lesssim\alpha<-0.5$ and the low-frequency spectral index of multi-component SED fits having $-3\lesssim\alpha_{\rm{low}}<+5$. Likewise, around half the sample of 60 pulsars detected by \citet{Murphy2017} were not well-described using a single power-law, with a number showing low-frequency flattening or turn-overs. As such, when searching our new all-sky catalogue for new candidate pulsars, we relaxed criterion (c) and instead searched for pulsar candidates by compactness and polarisation fraction.

Among our 484 nominally extragalactic sources, we found four additional candidates which are not visibly spatially extended\footnote{At the resolution of GLEAM or ancillary higher-resolution surveys (where available, the TGSS-ADR1, SUMSS and NVSS).}, exhibit a polarisation fraction $\Pi_{200\,\rm{MHz}} \geq10\%$, and do not have a likely IR counterpart in AllWISE. We note that the pulsar candidate presented in \citetalias{Riseley2018}, GLEAM~J134038$-$340234, is not detected in this all-sky catalogue. This suggests that this source is unlikely to be a true pulsar, as pulsars are expected to be Faraday-thin and would not depolarise significantly with the shift in reference frequency between \citetalias{Riseley2018} ($\nu_{\rm{ref}}=216$\,MHz) and this work ($\nu_{\rm{ref}}=200$\,MHz).

\begin{sidewaystable}
\centering
\footnotesize
\caption{New pulsar candidates identified from our nominally extragalactic source population, selected according to compactness and polarisation fraction. \label{tab:psrcand}}
\begin{tabular}{crrcccccc}
%\hline
Source Name & $l$ & $b$ & $S_{200\,\rm{MHz}}$ & $P_{200\,\rm{MHz}}$ & $\Pi_{200\,\rm{MHz}}$ & RM & $\alpha$ \\
 & & & [Jy] & [Jy] &  & [rad m$^{-2}$] & \\
\hline
GLEAM~J020549$-$791335 & 298.64384 & $-37.26391$ & $0.203\pm0.005$ & $0.044\pm0.024$ & $0.215\pm0.118$ & $-36.44\pm4.93$ & $-0.76\pm0.03$ \\
GLEAM~J030656$+$163144 & 163.87010 & $-35.36161$ & $0.221\pm0.007$ & $0.038\pm0.009$ & $0.173\pm0.040$ & $-7.71\pm1.95$ & $-0.76\pm0.02$ \\
GLEAM~J210917$-$720828 & 321.17232 & $-36.04911$ & $0.091\pm0.006$ & $0.050\pm0.010$ & $0.544\pm0.116$ & $+24.01\pm0.41$ & $-0.80\pm0.03$ \\
TGSSADR~J230010.0$+$184537 & 89.44966 & $-36.77971$ & $0.270\pm0.020$ & $0.043\pm0.008$ & $0.159\pm0.032$ & $-63.70\pm0.17$ & $-0.67\pm0.06$ \\
\hline
\end{tabular}
\end{sidewaystable}

We present these pulsar candidates in Table~\ref{tab:psrcand}. For the spectral index values listed in Table~\ref{tab:psrcand} we also sourced ancillary measurements from various radio surveys. As well as the VLSSr, TGSS-ADR1, TXS and NVSS catalogues, measurements were also found in catalogues from VLSSr the VLA's Faint Images of the Radio Sky at Twenty-cm survey \citep[FIRST;][]{Becker1994} and the Westerbork in the Southern Hemisphere survey \citep[WISH;][]{DeBreuck2002}. Flux density measurements from these surveys (where available) were combined with the GLEAM catalogue measurements and used to derive a single-component power-law fit, with the uncertainty region explored using a Monte-Carlo routine employed as part of the \verb|EMCEE| package \citep{emcee}. The resulting SEDs are presented in Figure~A3, except for TGSSADR~J230010.0$+$184537, where the SED is already shown in Figure~A2 (see Appendix~\ref{sec:appendix_a}).

From Table~\ref{tab:psrcand} (as well as Figure~A3) our candidate pulsars exhibit fairly typical synchrotron spectra, with $-0.8 < \alpha < -0.6$. While the GLEAM measurements for some of the candidates exhibit significant scatter, none of the spectra show clear signs of variability, or of multiple components. Time-domain observations with the MWA's VCS, for example, would be required to determine whether these sources are true pulsars, or simply strongly-polarised compact extragalactic radio sources.

\section{CONCLUSIONS AND OUTLOOK}
\subsection{Conclusions}
In this paper, we have presented the all-sky results from the POlarised GLEAM Survey (POGS). We have catalogued the low-frequency polarised radio source population, applying the RM synthesis technique to the GLEAM survey data, covering 25,489 square degrees of sky between Declination $+30\degree$ and $-82\degree$. We have detected a total of 517 radio sources, of which 33 are known radio pulsars and the remaining 484 are nominally extragalactic in origin. We have reported the bulk properties of our two catalogues at a reference frequency of 200\,MHz.

Our extragalactic catalogue, POGS ExGal, contains sources with linearly-polarised flux densities between 9.9\,mJy and 1.1\,Jy. All sources in POGS ExGal have RMs between $-328.07$~rad~m$^{-2}$ and $+279.62$~rad~m$^{-2}$. We find that our RMs are largely consistent with previous RM catalogues at higher frequencies. We determine RMs for these sources that are typically one or two orders of magnitude more precise than previous studies, with a mean and worst-case uncertainty of $0.38$~rad~m$^{-2}$ and $10.65$~rad~m$^{-2}$ respectively. Our results suggest that the dominant component of the RM is contributed by the Galactic foreground, although there is sufficient discrepancy that some extragalactic RM contribution must also be present. We have compared the bulk polarisation properties of our sources with the 1.4\,GHz polarised source population, finding that sources depolarise by about 55\% between 1.4\,GHz and 200\,MHz.

We find that the population of extragalactic radio sources shows significantly increasing fractional polarisation with decreasing Stokes $I$ flux density. From our sample, fainter Stokes $I$ sources ($S<0.5$\,Jy) tend to have a fractional polarisation that is $\sim2.6$ higher than brighter Stokes $I$ sources ($0.5<S<3$\,Jy). While this is consistent with some previous studies at higher frequencies, we are naturally biased toward fainter Stokes $I$ sources with higher fractional polarisation. 

We identify ten `polarised doubles', i.e. extended radio galaxies where a physical pair of polarised sources are detected, associated with opposing radio lobes. All show statistically significant RM variations between physically-related emission components. Our catalogue also contains 14 sources with large RMs, which we define as $|{\rm{RM}}| > 100$~rad~m$^{-2}$. The majority of these lie along lines of sight that pass through ionised Galactic foregrounds, visible in H$\alpha$ emission.

For our extragalactic source population, we observe no significant evolution in residual RM (which can be used to probe extragalactic magnetic fields) as a function of redshift. This has been seen previously at higher frequencies, but is seen here for the first time at low frequencies, which suggests that the observed Faraday rotation is occurring external to the radio sources. We also see an anticorrelation between RRM and polarisation fraction, which we attribute to a depolarisation mechanism. Given our long observing wavelength and our moderate resolution of around 3 to 7 arcminutes, we suggest that the responsible mechanism is beam depolarisation due to small-scale variations in the Galactic RM that occur within our beam element.

Among our nominally extragalactic radio source population, we find four sources that are compact, exhibit a high polarisation fraction $(\Pi \geq 10\%)$ and do not have a clear infrared host; these we identify as new pulsar candidates.

Our known-pulsar catalogue, POGS PsrCat, contains pulsars with linearly-polarised flux density measurements between 17\,mJy and 1.7\,Jy. Our pulsar RMs span the range $-185.99$~rad~m$^{-2}$ to $+150.74$~rad~m$^{-2}$. We find that our RMs are broadly consistent with known values, with a typical $\sim40\%$ improvement in the RM precision compared to previous measurements: our mean and worst-case uncertainties are $0.22$~rad~m$^{-2}$ and $2.84$~rad~m$^{-2}$, respectively. There are ten pulsars for which we make the first image-plane detection at low frequencies, and seven pulsars for which we determine the first RMs.

Our image-domain search for pulsars has yielded RMs and fractional polarisations that are broadly consistent with previous time-domain, beam-formed, studies, although we note that we likely suffer from strong scintillation. Our study also demonstrates that image-domain searches retain sensitivity to significantly-dispersed pulsars, as we find RMs for pulsars with DMs up to a factor $\sim2.5$ larger than previous time-domain beam-formed data.

\subsection{Further Work}
While this paper represents the final catalogue from our all-sky linear polarisation survey with the Phase I MWA, there are a number of novel aspects of the linearly-polarised source population that are left to explore. 

Foremost among these is the low-frequency linearly-polarised source counts, which remain entirely unexplored. However, due to our non-standard source identification and verification method, the completeness of our catalogue is non-trivial to establish, and will require injection of 3D source models into RM spectra---a novel adaptation of standard methods used in continuum source completeness evaluation. Such work is beyond the scope of this paper, but crucial for establishing the completeness, which is in turn key to probing the differential source counts.

Looking forward, polarisation work with the Phase II MWA will build on our work. The factor $\sim2$ improvement in resolution achievable with the extended configuration (up to $\sim6$\,km; \citealt{Wayth2018}) will provide a huge step forward for low-frequency polarimetry by significantly reducing beam depolarisation. Not only should this yield an increased number of detections across the sky, but a direct comparison of Phase I and Phase II MWA polarimetry using the same sources could provide insight into the scale size of Galactic foreground RM fluctuations. Both our work and Phase II MWA polarimetry also provide a crucial step in `filling in' the gap in the Southern RM sky.

\begin{acknowledgements}
This work makes use of the Murchison Radioastronomy Observatory, operated by CSIRO. We acknowledge the Wajarri Yamatji people as the traditional owners of the Observatory site. Support for the operation of the MWA is provided by the Australian Government (NCRIS) under a contract to Curtin University, administered by Astronomy Australia Limited. This work was supported by resources provided by the Pawsey Supercomputing Centre, with funding from the Australian Government and the Government of Western Australia.  

CJR acknowledges financial support from the ERC Starting Grant ``DRANOEL'', number 714245. The Dunlap Institute is funded through an endowment established by the David Dunlap family and the University of Toronto. B.M.G. acknowledges the support of the Natural Sciences and Engineering Research Council of Canada (NSERC) through grant RGPIN-2015-05948, and of the Canada Research Chairs program. CSA is a Jansky Fellow of the National Radio Astronomy Observatory. NHW is supported by an Australian Research Council Future Fellowship (project number FT190100231) funded by the Australian Government. We acknowledge the International Centre for Radio Astronomy Research (ICRAR), which is a joint venture between Curtin University and The University of Western Australia, funded by the Western Australian State government. The financial assistance of the South African Radio Astronomy Observatory (SARAO) towards this research is hereby acknowledged (\url{www.ska.ac.za}). We thank Phil Edwards for helpful comments during the internal ATNF review process, and we thank our anonymous referee for their supportive and constructive feedback during peer review.

This work has made use of S-band Polarisation All Sky Survey (S-PASS) data. All NVAS images used in this work were produced as part of the NRAO VLA Archive Survey, (c) AUI/NRAO. This research has made use of NASA's Astrophysics Data System (ADS) as well as the VizieR catalogue access tool, CDS, Strasbourg, France. This research has made use of the NASA/IPAC Extragalactic Database (NED), which is funded by the National Aeronautics and Space Administration and operated by the California Institute of Technology. Additionally, this research made use of the Tool for OPerations on Catalogues And Tables (TOPCAT) software \citep{TOPCAT}. We also used ionospheric TEC maps produced by the Centre for Orbital Determination in Europe (CODE; \url{http://aiuws.unibe.ch/ionosphere/}). We acknowledge the use of NASA’s SkyView service (\url{http://skyview.gsfc.nasa.gov}), located at the NASA Goddard Space Flight Center. Finally, we wish to acknowledge the developers of the following python packages, which were used extensively during this project: \texttt{aplpy} \citep{Robitaille2012}, \texttt{astropy} \citep{Astropy2013}, \texttt{lmfit} \citep{Newville2017}, \texttt{matplotlib} \citep{Hunter2007}, \texttt{numpy} \citep{Numpy2011} and \texttt{scipy} \citep{Jones2001}.

\end{acknowledgements}

\bibliographystyle{pasa-mnras}
\bibliography{POGS-II}

\appendix
\counterwithin{figure}{section}

\section{Spectral Energy Distribution Fits}\label{sec:appendix_a}
In this section, we show SED fits used to derive 200 MHz Stokes I flux densities for some sources in our catalogues. Figure~A1 shows the SEDs for GLEAM sources which have large fractional uncertainty in`\verb|int_flux_fit_200|' from \cite{HurleyWalker2017}. Figure~A2 shows the SEDs for TGSS-ADR1 sources. Figure~A3 shows the SEDs for new pulsar candidates identified in this work.

\section{Polarised doubles}\label{sec:appendix_b}
Two `polarised doubles'---where we define a `polarised double' as a pair of physically-associated polarised sources---were previously presented in Figure~\ref{fig:exgal_doubles}. We present the remaining eight polarised doubles in POGS ExGal in Figure~B1.

\section{Pulsar RM spectra}\label{sec:appendix_c}
We present the RM spectra for the 33 known pulsars identified in POGS PsrCat in Figure~C1. As with RM spectra shown elsewhere in this paper, the source spectrum is shown in black and the off-source foreground spectrum is shown in red. The blue $7\sigma$ level is shown by the blue dot-dashed horizontal line, and the fitted peak shown as a vertical green dashed line. The polarised flux density is shown in mJy beam$^{-1}$.

\section{Extragalactic source RM spectra}\label{sec:appendix_d}
Figure~D1 presents the RM spectra of all POGS ExGal sources, aside from the ten `polarised doubles'. As with previous RM spectra, the source spectrum is shown in black and the off-source foreground spectrum is shown in red. The $7\sigma$ level is shown by the blue dot-dashed horizontal line, and the fitted peak shown as a vertical green dashed line. The polarised flux density is shown in mJy beam$^{-1}$.

\end{document}